\documentclass[twocolumn]{aastex631}

\usepackage[whole]{bxcjkjatype}

\usepackage[whole]{bxcjkjatype} 
\usepackage[utf8]{inputenc}

\accepted{to ApJ on February 25, 2025}

\shorttitle{R-JET: a postprocessing code for radiative transport in jets}
\shortauthors{Hirotani et al.}
\graphicspath{{./}{figures/}}

\begin{document}

\title{{\tt R-JET}: 
A postprocessing code for radiative transport in relativistic jets
}

\correspondingauthor{Kouichi Hirotani, Hsien Shang}
\email{hirotani@asiaa.sinica.edu,shang@asiaa.sinica.edu}

\author[0000-0002-2472-9002]{Kouichi Hirotani}
\affiliation{Institute of Astronomy and Astrophysics, Academia Sinica, Taipei 106216, Taiwan}

\author[0000-0001-8385-9838]{Hsien Shang （尚賢）}
\affiliation{Institute of Astronomy and Astrophysics, Academia Sinica, Taipei 106216, Taiwan}

\author[0000-0001-5557-5387]{Ruben Krasnopolsky}
\affiliation{Institute of Astronomy and Astrophysics, Academia Sinica, Taipei 106216, Taiwan}

\author[0000-0001-6031-7040]{Kenichi Nishikawa}
\affiliation{Department of Physics, Chemistry and Mathematics, 
Alabama A\&M University, Huntsville, AL 35811, USA
}

\begin{abstract}
We describe a post-processing radiative transport code for computing the spectra, the coreshift, and the surface-brightness distribution of special relativistic jets with arbitrary optical thickness. 
The jet consists of an electron-positron pair plasma
and an electron-proton normal plasma.
Electrons and positrons are relativistic
and composed of thermal and nonthermal components,
while protons are non-relativistic and non-radiating.
The fraction of a pair plasma, 
as well as the fraction of a nonthermal component
can be arbitrarily chosen.
Only the synchrotron process is considered for emission and absorption  
when the radiative-transfer equation is integrated along our lines of sight.
We describe a suite of test problems, 
and confirm the frequency dependence of the coreshift
in the Konigl jet model,
when the plasma is composed of nonthermal component alone.
Finally, we illustrate the capabilities of the code with model calculations,
demonstrating that the jet will exhibit a limb-brightened structure
in general when it is energized by the rotational energy of the black hole.
It is also demonstrated that such limb-brightened jets show
a ring-like structure in the brightness map
when we observe the jet launching region nearly face-on.
\end{abstract}


\section{Introduction}
\label{sec:intro}
At the center of active galaxies, 
accreting supermassive black holes (SMBHs) emit powerful jets, 
seen as tightly collimated beams of relativistic plasmas.
Understanding the formation and initial acceleration of 
these jets is still a major unsolved question in modern astrophysics.
In radio frequencies,
high-resolution Very-Long-Baseline Interferometry (VLBI)
imaging of inner jet regions provides 
the best constraints for the jet formation theories.
The recent advent of The Event Horizon Telescope (EHT),
the global short-millimeter VLBI project,
has allowed for the observation of M87*
(i.e., the center of radio galaxy M87)
at spatial scales comparable to the event horizon.
The jet of M87 was resolved down to $40\mu$as,
or equivalently to $5.5 R_{\rm S}$ at 230~GHz
\citep{Doeleman:2012:Sci},
where $R_{\rm S}$ refers to the Schwarzschild radius.
In addition, the measurement of the coreshift (due to the opacity effect) 
suggest that the radio core is located around $10 R_{\rm S}$
in de-projected distance from the black hole (BH)
\citep{hada11}.
What is more, a limb-brightened structure is found in the M87 jet
within the central $90 R_{\rm S}$ in de-projected distance
at 86~GHz \citep{hada16}.

From theoretical point of view, 
these relativistic flows
are energized either by the extraction of the BH's rotational energy
via the magnetic fields threading the event horizon
\citep{bla77} 
or by the extraction of the accretion flow's rotational energy
via the magnetic fields threading the accretion disk
\citep{blandford:1982MNRAS}.
In the recent two decades, 
general relativistic (GR) magnetohydrodynamic (MHD) simulations
successfully demonstrated that a Poynting-dominated jet can be driven by
the former process, the so-called \lq\lq Blandford-Znajek (BZ) process''
\citep{Koide:2002:Sci,mckinney:2004ApJ,komissarov:2005MNRAS,
Tchekhovskoy:2011:MNRAS,Qian:2018ApJ}.
The BZ process takes place 
when the horizon-penetrating magnetic field is dragged
in the rotational direction of the BH by space-time dragging,
and when a meridional current is flowing in the ergo region,
where the BH's rotational energy is stored.
The resultant Lorentz force
exerts a counter torque on the horizon,
extracting the BH's rotational energy and angular momentum
electromagnetically,
and carrying them to large distances in the form of torsional Alfven waves.
Subsequently, applying the GR particle-in-cell (PIC) technique
to magnetically dominated BH magnetospheres,
it is confirmed that the BZ process still works
even when the Ohm's law, and hence the MHD approximation breaks down
in a collisionless plasma
\citep{Parfrey:2019:PhRvL,Chen:2020:ApJ,Kisaka:2020:ApJ,Crinquand:2021:AA,
Bransgrove:2021:PhRvL,Hirotani:2023:ApJ}.

It is possible that such extracted electromagnetic energy
is converted into the kinetic and internal energies 
of the jet plasmas,
and eventually dissipated as radiation 
via the synchrotron and the inverse-Compton (IC) processes 
in the jet downstream.
In this context, 
we propose a method to infer the plasma density in a stationary jet,
connecting the Poynting flux due to the BZ process
to the kinetic flux in the jet downstream. 
The conversion efficiency from Poynting to kinetic energies
is parameterized by the magnetization parameter $\sigma$,
which evolves as a function of the distance $r$ from the BH.
At the same time, the toroidal component of the magnetic field
can also be parameterized by $\sigma$,
while the poloildal component by the flow-line geometry. 
Accordingly, the plasma density and the magnetic field strength
can be inferred at each position in the jet,
once the evolution of $\sigma$ and $\Gamma$
(the bulk Lorentz factor) is specified.
We then compute the emission and absorption coefficients
at each position of the jet,
and integrate the radiative transfer equation
along the lines of sight in our
post-processing radiative transport code, {\tt R-JET}.
In this code, we concentrate on the synchrotron process
as the radiative process in jets,
neglecting the IC processes in the current version.
The {\tt R-JET} code works for arbitrary ratios of 
pair and normal plasmas,
arbitrary ratios of thermal and nonthermal plasmas,
arbitrary flow lines of axisymmetric jet, and
arbitrary evolution of $\sigma$ and $\Gamma$ with $r$.

We start with analytically describing our parabolic jet model
in \S~\ref{sec:parabolic_jet},
focusing on how to connect the Poynting-dominated jet at the jet base
and the kinetic-dominated jet in the downstream.
In \S~\ref{sec:verification},
we check the code by comparing the numerical results
with analytic solutions.
Then in \S~\ref{sec:implications},
we apply the 
the {\tt R-JET} code to typical BH and jet parameters
and demonstrate that 
a limb-brightened structure (as found in the M87 jet)
is a general property of
BH jets when the BH's rotational energy is efficiently extracted
via the BZ process.
It is also demonstrated that such limb-brightened jets exhibit 
a ring-like structure 
\citep[as found in M87* at 86~GHz,][]{Lu:2023:Natur} 
when relativistic plasmas are supplied in jet
at a certain distance from the BH.
We summarize this work in \S~\ref{sec:summary}.

\section{Parabolic jet model}
\label{sec:parabolic_jet}
In the present paper, we consider stationary and axisymmetric jet.
We assume that the jet is 
composed of an electron-positron pair plasma
and an electron-proton normal plasma.
For electrons and positrons, 
which are referred to as leptons in the present paper,
we consider both thermal and nonthermal components.
We assume protons are non-relativistic in the jet co-moving frame.

\subsection{Poynting flux near the Black Hole}
\label{sec:BZflux}
To model the angle-dependent Poynting flux in the jet-launching region,
we follow the method described in \S~2 of 
\citet[][hereafter H24]{Hirotani:2024:ApJ}.
We model the jet flow lines with a parabolic-like geometry,
adopting the magnetic-flux function
\citep{2009ApJ...697.1164B,Takahashi:2018:ApJ},
\begin{equation}
  A_\varphi
    = A_{\rm max} 
    \left( \frac{r}{R_{\rm S}} \right)^q
    \left( 1 - \vert \cos\theta \vert \right),
  \label{eq:A3}
\end{equation}
where 
$R_{\rm S}= 2 G M / c^2 = 2M$ denotes the Schwarzschild radius
in geometrized unit (i.e., $c=G=1$), 
and $\theta$ does the colatitude;
$c$ and $G$ refer to the speed of light and the gravitational constant,
respectively.
The flow line becomes parabolic when $q=1$,
and conical when $q=0$.
For the M87 jet, a quasi-parabolic flow-line geometry, $q=0.75$, 
is observationally suggested within the Bondi radius
\citep{Asada:2012:ApJL}.

Differentiating equation~(\ref{eq:A3}) with respect to $r$ and $\theta$, 
we obtain the meridional and radial components of the magnetic field, 
respectively (Appendix).
Accordingly, the strength of the magnetic field 
in the poloidal plane is obtained 
at position ($r$,$\theta$) as 
\begin{equation}
  B_{\rm p}
  = B_{\rm p,0}
    \left( \frac{r}{R_{\rm S}} \right)^{q-2}
    \sqrt{ 1 + q^2 \left( \frac{1 - \cos\vert\theta\vert}
                               {\sin\theta}
                   \right)^2
                   H^2
          },
  \label{eq:Bp}
\end{equation}
where 
\begin{equation}
  H \equiv 1+\frac{r}{q} \frac{dq}{dr} 
             \ln\left(\frac{r}{R_{\rm S}}\right).
\end{equation}
The magnetic field strength at $r= R_{\rm S}$ becomes
$B_{\rm p} \approx B_{\rm p,0}$ near the pole ($\sin\theta \approx 0$),
and becomes $B_{\rm p} \approx B_{\rm p,0} \sqrt{1+q^2}$
near the equator ($\theta \approx \pi/2$).
The variation of $q$ as a function of $r$ is taken into account
in $B_{\rm p}$ through the function 
$H=H(r)$.

In general, $q$ is a function of $A_\varphi$ and $r$.
However, in the present paper, we assume that $q$ is constant
in the entire magnetosphere for simplicity.

Let us consider the Poynting flux, which is given by
\begin{equation}
  T^r{}_t 
  = \frac{1}{4\pi} 
    \left[ F^{\mu\alpha} F_{\alpha t} 
          +\frac{1}{4} g^{\mu}_t F^{\alpha\beta}F_{\alpha\beta}
    \right]
  = \frac{1}{4\pi} F^{r\theta}F_{\theta t},
  \label{eq:Poynting_flux}
\end{equation}
where $F_{\mu\nu}$ denotes (a covariant component of) 
the Faraday tensor.
The Greek indices $\alpha$ and $\beta$ run over $0$, $1$, $2$, and $3$.
In the second equality, we put 
$F_{\varphi t}=0$ assuming stationary and axisymmetric magnetosphere.

In the Boyer-Lindquist coordinates, we obtain
the covariant component of the magnetic-field four vector:
\begin{equation}
  B_\varphi= -\frac{\rho_{\rm w}^2}{\sqrt{-g}} F_{r\theta}
  = -\Sigma \sin\theta F^{r\theta},
  \label{eq:B_phi}
\end{equation}
where 
$\rho_{\rm w}$ denotes the distance from the rotation axis,
$\sqrt{-g}= \Sigma \sin\theta$, and
$\Sigma \equiv r^2 + a^2 \cos^2\theta$; 
$r$ and $\theta$ denotes the radial coordinate and the colatitude
in the polar coordinates 
(or in this case, in the Boyer-Lindquist coordinates).
The BH's spin parameter $a$ becomes
$a=0$ (or $a=M$) for a non-rotating (or an extremely rotating) BH;
$M$ refers to the BH mass 
(or the gravitational radius in the geometrical unit).
In general, the magnetic-field four vector is defined by
$B_\mu \equiv {}^\ast\! F_{\mu t}$,
where 
${}^\ast\! F_{\mu\nu}$ denotes the Maxwell tensor
(i.e., the dual of the Faraday tensor).
In the AGN-rest frame, the physical strength of 
the toroidal magnetic field is given by the orthonormal component,
\begin{equation}
 B_{\hat\varphi}
  = \frac{B_\varphi}{\sqrt{g_{\varphi\varphi}}},
  \label{eq:B_toroidal}
\end{equation}
where $\sqrt{g_{\varphi\varphi}} = r \sin\theta$
denotes the distance from the rotation axis in a flat spacetime.
Although equation~(\ref{eq:B_toroidal}) is also valid in GR
as the toroidal component measured by a Zero-Angular-Momentum Observer
(ZAMO) in the Kerr spacetime, 
we employ the special relativistic formalism 
in the current version of the {\tt R-JET} code, 
neglecting GR corrections.

In ideal MHD, the frozen-in condition gives
the meridional electric field,
$F_{\theta t}= -\Omega_{\rm F} F_{\theta\phi}$,
where $\Omega_{\rm F}$ 
denotes the angular frequency of rotating magnetic field lines.
The radial component of the magnetic field is obtained by
\begin{equation}
  \tilde{B}^r= \frac{F_{\theta\phi}}{\sqrt{-g}}
  = \frac{A_{\rm max}}{\Sigma} \left(\frac{r}{r_{\rm H}} \right)^q,
  \label{eq:Br}
\end{equation}
where equation~(\ref{eq:A3}) is used in the second equality.
Thus, at the horizon, we obtain 
$B_{\rm p} \approx \tilde{B}^r= A_{\rm max}/\Sigma$.

We can constrain the meridional distribution of 
$F^{r\theta}= -B_\varphi / \sqrt{-g} $,
using MHD simulations in the literature.
For an enclosed electric current $I(A_\varphi)$
in the poloidal plane, we obtain
\begin{equation}
  B_\varphi \approx -I(A_\varphi) / 2 \pi.
  \label{eq:Bphi}
\end{equation}
By GRMHD simulations,
\citet{Tchekhovskoy:2010:ApJ} derived the current density
\begin{equation}
  I(A_\varphi) 
  \approx 6 \sin\left[ \frac{\pi}{2} (1-\cos\theta) \right] 
          (\omega_{\rm H}-\Omega_{\rm F}) B_{\rm p}
\end{equation}
in the poloidal plane, 
where $\Omega_{\rm F} \approx 0.5 \omega_{\rm H}$ is assumed.
Note that $I(A_\varphi)$ is proportional to the Goldreich-Julian
current density \citep{GJ69,bes92,hiro06}, 
$(\omega_{\rm H}-\Omega_{\rm F}) B_{\rm p} / 2 \pi$.
We thus obtain
\begin{equation}
  B_\varphi
  \approx -\frac{3}{\pi}
    \sin\left[ \frac{\pi}{2} (1-\cos\theta) \right]
    (\omega_{\rm H}-\Omega_{\rm F}) B_{\rm p}.
  \label{eq:B_phi2}
\end{equation}
Combining equations~(\ref{eq:Poynting_flux}),
(\ref{eq:B_phi}), (\ref{eq:Br}), and (\ref{eq:B_phi2}),
we obtain the angle-dependent BZ flux,
\begin{equation}
  T^r{}_t
  \approx \frac{3}{4\pi^2} 
    \sin\left[ \frac{\pi}{2} (1-\cos\theta) \right] 
    \Omega_{\rm F} ( \omega_{\rm H}-\omega_{\rm H} ) 
    B_{\rm p,H}^2,
  \label{eq:Poynting_2}
\end{equation}
where $B_{\rm p,H} \equiv B_{\rm p}(r=r_{\rm H})$.
It follows from its $\theta$ dependence
that the BH's rotational energy
is preferentially extracted along the magnetic field lines
threading the horizon in the lower latitudes,
$\theta \sim \pi/2$.

\subsection{Kinetic flux inferred by the BZ flux}
\label{sec:jet_kinetic}
In this subsection,
we connect the BZ flux inferred near the horizon
(\S~\ref{sec:BZflux})
with the kinetic flux in the jet downstream.
To this end, we describe the kinetic flux
in terms of bulk Lorentz factor and the co-moving energy density.

Considering that the jet plasma is composed of 
an electron-positron pair plasma and 
an electron-proton normal plasma,
we can describe the proper number density of pair-origin electrons by
\begin{equation}
  n_{\ast,{\rm e}}^{\rm pair}
  = f_{\rm p} n_{\ast,{\rm e}}^{\rm tot},
\end{equation}
where $n_{\ast,{\rm e}}^{\rm tot}$ represents 
the density of electrons of both origins, 
and $f_{\rm p}$ does the fraction of pair contribution in number;
$n_\ast$ denotes the number density in the jet co-moving frame.
The proper number density of protons is parameterized by
$n_{\ast,{\rm p}}= ( 1 - f_{\rm p} ) n_{\ast,{\rm e}}^{\rm tot}$,
where the normal plasma is assumed to be composed of pure hydrogen.
A pure {\it pair} plasma is obtained by $f_{\rm p}= 1$, whereas
a pure {\it normal} plasma by $f_{\rm p}= 0$.

What is more, we assume that the leptons consist of
thermal and nonthermal components.
Introducing a nonthermal fraction $w_{\rm nt}$,
we can write down the nonthermal and thermal electron densities as
\begin{equation}
  n_{\ast,{\rm e}}^{\rm nt}
  = w_{\rm nt} n_{\ast,{\rm e}}^{\rm tot},
  \label{eq:n_nt}
\end{equation}
and
\begin{equation}
  n_{\ast,{\rm e}}^{\rm th}
  = (1-w_{\rm nt}) n_{\ast,{\rm e}}^{\rm tot}
  \label{eq:n_th}
\end{equation}
respectively.

In the BH-rest frame,
the kinetic flux can be expressed as
\begin{equation}
  F_{\rm kin}
  = \beta c \Gamma (\Gamma-1) n_{\ast,{\rm e}}^{\rm tot} U,
  \label{eq:F_kin}
\end{equation}
where $\beta c$ denotes the fluid velocity,
$\Gamma \equiv 1 / \sqrt{1-\beta^2}$ the bulk Lorentz factor,
and $U$ the fluid mass per electron in the jet co-moving frame.  
The factor $\Gamma$ appears due to the Lorentz contraction,
and another factor $\Gamma-1$ expresses
the kinetic contribution in energies
(i.e., total energy minus rest-mass energy).

Using $f_{\rm p}$ and $w_{\rm th}$,
we can express $U$ as the summation of 
leptonic and hadronic contributions,
\begin{eqnarray}
  U
  &\equiv& \left[ \frac32 \Theta_{\rm e} ( 1 - w_{\rm nt} )
                 + \langle \gamma \rangle w_{\rm nt} 
           \right] \cdot (1+f_{\rm p}) m_{\rm e} c^2
  \nonumber\\
  && + ( 1 - f_{\rm p} ) m_{\rm p} c^2.
  \label{eq:def_U}
\end{eqnarray}
where $\Theta_{\rm e} \equiv kT_{\rm e} / m_{\rm e} c^2$ denotes
the dimensionless temperature of thermal leptons,
$\langle \gamma \rangle$ 
the mean Lorentz factor of randomly moving nonthermal leptons,
$m_{\rm e}c^2$ the electron's rest-mass energy, and
$m_{\rm p}c^2$ the proton's rest-mass energy.
A factor $1+f_{\rm p}$ appears 
in the first line 
due to the contribution of positrons in addition to electrons.
We assume here that protons are non-relativistic.
The existence of a normal plasma (i.e., $f_{\rm p}<1$) 
becomes important in the present analysis
when the second line dominates the first line
through the heavier proton mass.

Neglecting the energy dissipation in the jet
\citep{Celotti:1993:MNRAS:264},
we find that the summation of electromagnetic and kinetic energies
is conserved along each flux tube.
We thus obtain the kinetic-energy flux
\begin{equation}
  F_{\rm kin}(r,\theta)
  = \frac{1}{1+\sigma}
    \frac{B_{\rm p}(r,\theta)}{B_{\rm p,0}} 
    F_{\rm BZ,0}
  \label{eq:F_kin4}
\end{equation}
at position ($r$,$\theta$) in the jet,
where $\sigma$ denotes the magnetization parameter,
$B_{\rm p}(r,\theta)$ the poloidal magnetic field strength
at distance $r$ from the BH and at colatitude $\theta$ in the jet.
The two constants, $F_{\rm BZ,0}$ and $B_{\rm p,0}$, denote
the Poynting flux and the poloidal-magnetic-field strength,
respectively, at $r= R_{\rm S}$.
The factor $B_{\rm p} / B_{\rm p,0}$ 
shows how the flow-line cross section increases outwards
as the magnetic flux tube expands.

Note that the toroidal component of the magnetic field evolves by
\begin{equation}
  B_{\hat\varphi}= \frac{\sigma}{1+\sigma} B_{\hat\varphi}(R_{\rm S}).
\end{equation}
In this case, the Poynting flux evolves as
\begin{equation}
  F_{\rm EM}(r,A_\varphi)
  = \frac{\sigma}{1+\sigma} 
    \frac{B_{\rm p}(r,A_\varphi)}{B_{\rm p}(R_{\rm S},A_\varphi)}
    F_{\rm BZ}(R_{\rm S},A_\varphi).
\end{equation}
Accordingly, the total energy per magnetic flux tube,
\begin{equation}
  \frac{F_{\rm EM}+F_{\rm kin}}{B_{\rm p}}
  = \frac{F_{\rm BZ}}{B_{\rm p}(R_{\rm S})}
\end{equation}
is conserved along each flow line.

If the magnetic field geometry is conical near the BH,
we obtain $F_{\rm BZ,0}= T^r{}_t$ (eq.~[\ref{eq:Poynting_2}]).
Since the magnetic field lines are more or less radial
at the horizon due to the plasma inertia and the causality
at the horizon,
we evaluate $F_{\rm BZ,0}$ by equation~(\ref{eq:Poynting_2}) 
even when $q \ne 0$ at the horizon for simplicity.
The normalization factor, $A_{\rm max}$
is related to $B_{\rm p,0}$ by
$A_{\rm max}= B_{\rm p,0} R_{\rm S}^2$.

Equations~(\ref{eq:F_kin}) and (\ref{eq:F_kin4}) show that
electron density, $n_{\ast,{\rm e}}^{\rm tot}$, 
can be computed if $\Gamma$, $\sigma$, $f_{\rm p}$, and $w_{\rm nt}$
are specified at each position of the jet.
In general, $\Gamma$ can be observationally constrained 
by the super-luminal motion of individual jet component
and the observer's viewing angle.
We could constrain $w_{\rm nt}= w_{\rm nt}(r,A_\varphi)$ 
by comparing the predicted SED with observations.
To constrain the functional form of $\sigma=\sigma(r,A_\varphi)$,
we compare the predicted SED and the coreshift with 
VLBI observations.
As for $f_{\rm p}= f_{\rm p}(r,A_\varphi)$,
it is not straightforward to constrain the functional form.
However, we will give some hints on this issue
in \S~\ref{sec:impl_content}.

\subsection{Lepton energy distribution}
\label{sec:jet_energy}
We assume that the leptons are re-accelerated
only within the altitude $r < r_1$,
and adopt $r_1= 800 R_{\rm S}$.
As long as $r_1$ is large enough,
the actual value of $r_1$ does not affect the result,
because the synchrotron emission from large radii
little changes the SED, coreshift, and the brightness map.
Therefore, we consider that
$800 R_{\rm s}$ is a safe upper limit of $r_1$
in practice.

At $r<r_1$, we adopt the following form 
of the nonthermal lepton energy distribution,
\begin{equation}
  \frac{dn_{\ast,{\rm e}}^{\rm nt}}{d\gamma}
  = (p-1) n_{\ast,{\rm e}}^{\rm nt} \gamma^{-p},
  \label{eq:distr_nonth_1}
\end{equation}
where $p>1$ is assumed.
The nonthermal lepton density $n_{\ast,{\rm e}}^{\rm nt}$
is specified by $w_{\rm nt}$ (eq.~[\ref{eq:n_nt}]),
once $n_{\ast,{\rm e}}^{\rm tot}$
is constrained by equations~(\ref{eq:F_kin}), (\ref{eq:def_U}),
and (\ref{eq:F_kin4}).
We assume that the lower bound of the Lorentz factors, $\gamma$,
of randomly-moving, nonthermal leptons is 
$\gamma_{\rm min}=1$ throughout the jet.
For the upper bound of $\gamma$, we adopt
$\gamma_{\rm max}=10^5$.

At $r>r_1$, on the other hand, we must consider the evolution
of $\gamma_{\rm max}$ by {\it adiabatic} and {\it synchrotron} 
processes.
It is noteworthy that $\gamma_{\rm max}$
affects the SED through the turnover frequency,
although it does not contribute
in the integration of $dn_{\ast,{\rm e}}^{\rm nt}/d\gamma$
over $\gamma$
owing to the hard energy spectrum, $p>1$.

By {\it adiabatic} expansion,
the power-law index is unchanged
whereas the normalization decreases in the lepton energy distribution.
Thus, at $r>r_1$, we adopt the distribution function
\begin{equation}
  \frac{dn_{\ast,{\rm e}}^{\rm nt}}{d\gamma}
  = (p-1) n_{\ast,{\rm e}}^{\rm nt}(r_1)
    \left(\frac{B_{\rm p}}{B_{{\rm p},1}} \right)^{p-2} \gamma^{-p}
  \label{eq:distr_nonth}
\end{equation}
where $B_{{\rm p},1}$ denotes $B_{\rm p}$ at $r= r_1$.
The factor $B_{\rm p}/B_{{\rm p},1}$ indicates the
expansion factor of the fluid element.
Each lepton decreases its energy via adiabatic expansion; 
thus, $\gamma_{\rm max}$ evolves by
\begin{equation}
  \gamma_{\rm max}
  = \gamma_{\rm max,1} \cdot (B_{\rm p}/B_{{\rm p},1})^{\gamma_{\rm ad}-1},
  \label{eq:gamma_max_ad}
\end{equation}
where $\gamma_{\rm max,1} (= 10^5)$ denotes $\gamma_{\rm max}$ at $r=r_1$, 
and $\gamma_{\rm ad}$ denotes the adiabatic index.
Since nonthermal leptons are relativistic, 
we adopt $\gamma_{\rm ad}=4/3$.
Then equation~(\ref{eq:gamma_max_ad}) gives
$\gamma_{\rm max}$ at each $r$ when the adiabatic cooling dominates.
Let us denote this $\gamma_{\rm max}$ as $\gamma_{\rm max,ad}$.

By {\it synchrotron} radiation,
each lepton loses energy at the rate
\begin{equation}
  - m_{\rm e} c^2 \frac{d\gamma}{dt}
  = \frac43 \sigma_{\rm T} c \gamma^2 \frac{B^2}{8\pi},
\end{equation}
where $\sigma_{\rm T}$ denotes the Thomson cross section.
Thus, we compute the evolution of lepton Lorentz factors by
\begin{equation}
  d\gamma
  = -\frac{1}{6\pi} \frac{\sigma_{\rm T}}{m_{\rm e}c^2}
     B^2 \gamma^2 dr.
\end{equation}
Assuming the initial Lorentz factor is
$\gamma=\gamma_{\rm max,1}$ at $r=r_1$,
we can tabulate $\gamma_{\rm max}$ as a function of $r ( > r_1 )$
when synchrotron cooling dominates.
Let us denote this $\gamma_{\rm max}$ as $\gamma_{\rm max,syn}$.

Finally, we evaluate the maximum Lorentz factor by
\begin{equation}
  \gamma_{\rm max}
  = {\rm min}( \gamma_{\rm max,ad} , \gamma_{\rm max,syn})
  \label{eq:gamma_max}
\end{equation}
at $r > r_1$.
We use equation~(\ref{eq:distr_nonth}) to evaluate 
the energy distribution of nonthermal leptons,
and compute the emission and absorption coefficients at $r>r_1$.

For thermal pairs, we assume that they
have a semi-relativistic temperature. 
By adiabatic expansion, 
the lepton temperature $\Theta_{\rm e}$ evolves as
\begin{equation}
  \Theta_{\rm e}
  = \Theta_{\rm e,0} 
    \left( \frac{B_{\rm p}}{B_{\rm p,0}} \right)^{\gamma_{\rm ad}-1},
  \label{eq:temp}
\end{equation}
where $\Theta_{\rm e,0}$ denotes the temperature,
and $B_{\rm p,0}$ the poloidal magnetic field strength,
both at $r=R_{\rm S}$.
We set $\gamma_{\rm ad}=4/3$,
because thermal leptons contribute significantly
only within the relativistic regime.
We could instead set $\gamma_{\rm ad}=5/3$ when $\Theta_{\rm e} < 1$;
nevertheless, we could not find meaningful difference by this treatment,
because non-relativistic thermal leptons do not affect the results any way.
Note that a fluid element expands by a factor
$B_{\rm p,0}/B_{\rm p}$ in the jet compared at the jet base.

We assume an energy equipartition between
the energy density of (thermal+nonthermal) pairs
and that of the random magnetic field, 
$B_{\ast,{\rm ran}}^2/8\pi$.
For generality, we introduce a parameter $\kappa$,
and evaluate $B_{\ast,{\rm ran}}$ by
\begin{equation}
  \frac{B_{\ast,{\rm ran}}^2}{8\pi}
  = \kappa n_{\ast,{\rm e}}^{\rm tot} U,
  \label{eq:B_ran}
\end{equation}
where we put $\kappa= 1.0$ in the present paper.
Combining equation~(\ref{eq:B_ran}),
the definition of $\sigma$,
$\sigma \equiv F_{\rm EM} / F_{\rm kin}$, and 
$F_{\rm EM}=(c/4\pi) B_r B_{\hat\varphi}$,
we obtain
\begin{equation}
  B_{\ast,{\rm ran}}
  = \left( \frac{2\kappa}{\Gamma(\Gamma-1)}
           \frac{B_r B_{\hat\varphi}}{\sigma}
    \right)^{1/2}.
  \label{eq:B_ran_2}
\end{equation}
We use equation~(\ref{eq:B_ran_2}) to evaluate
the random magnetic field strength in the {\tt R-JET} code.
It is worth noting that in a kinetic-dominated jet,
$\sigma \ll 1$,
we obtain $B_{\hat\varphi} \propto \sigma$.
Accordingly, the detailed functional form of $\sigma$ 
on $r$ does not affect $B_{\ast,{\rm ran}}$.

On the other hand, for simplicity, 
we do not consider an energy equipartition for the ordered magnetic field.
It is noteworthy that such a large-scale magnetic field is produced
by the magnetospheric currents flowing both inside and outside the jet.
For instance, the Wald solution,
which corresponds to a GR extension of the cylindrical magnetic field
to the Kerr spacetime, is obtained when a ring current is flowing
at a large distance on the equatorial plane.
A parabolic magnetic field can be also produced by a toroidal current
flowing on the equatorial plane with a specific radial dependence.
The toroidal component of a magnetic field, $B_{\hat\varphi}$ is, 
on the other hand,
produced by the magnetospheric currents flowing in the poloidal plane.
Since it is out of the scope of the present paper
to infer the conduction currents in the jet,
we constrain $B_{\hat\varphi}$
with an assumed functional form of $\sigma(r)$ in the present framework,
instead of considering its energy balance with hot, relativistic plasmas.

The strength of the total magnetic field is evaluated by
$B= \sqrt{ B_{\rm p}^2 + B_{\hat\varphi}^2 + B_{\ast,{\rm ran}}^2 }$
at each position.

\subsection{Emission and absorption coefficients}
\label{sec:emi_abs_coeff}
Let us describe the emission and absorption coefficients
for thermal and nonthermal electron-positron pairs.
In this subsection (\S~\ref{sec:emi_abs_coeff}),
all the quantities are expressed in the jet co-moving frame,
although the asterisk ($\ast$) is omitted.

For thermal pairs, we assume that their energy distribution obeys 
the Maxwell-J$\ddot{\rm u}$ttner distribution.
We then obtain the emission coefficient
\citep{Leung:2011:ApJ,Wardzinski:2000:MNRAS},
\begin{eqnarray}
  &&
  j_\nu^{\rm (th)}
  = \frac{2\sqrt{2}\pi}{27}
    \frac{\Theta^2}{K_2(\Theta^{-1})}
    \frac{e^2}{c}
    \nu_{\rm L}
    n_\ast
  \nonumber \\
  &\times&
    X^{1/3} \left( X^{1/3} + 2^{11/12} \right)^2
    \exp\left( -X^{1/3} \right),
    \label{eq:emis_th}
\end{eqnarray}
where $X \equiv \nu / \nu_{\rm s}$,
$\nu_{\rm s} \equiv (2/9) \nu_{\rm L} \Theta^2 \sin\chi$,
$\nu_{\rm L} \equiv eB \sin\chi / ( 2\pi m_{\rm e} c)
 = 2.799 \times 10^6 B \sin\chi$;
$K_2$ denotes the modified Bessel function of the second kind 
of order $2$.
The pitch angle can be set as $\chi \sim 60^\circ$.
Here, $B$ denotes the magnetic field strength
and $\nu$ does the photon frequency,
both in the jet co-moving frame.

The absorption coefficient can be obtained by applying 
the Kirchhoff's law,
\begin{equation}
  \alpha_\nu^{\rm (th)}
  = \frac{j_\nu^{\rm (th)}}{B_\nu(T_{\rm e})},
  \label{eq:abso_th}
\end{equation}
where $B_\nu(T_{\rm e})$ denotes the Planck function.

For nonthermal pairs, we assume a power law energy distribution,
\begin{equation}
  \frac{d n_\ast}{d\gamma}
  = n_0 \gamma^{-p},
\end{equation}
where $\gamma$ denotes the Lorentz factor associated
with their random motion, and $p$ the power-law index.
Assuming that this power-law holds
within the Lorentz factor range,
$\gamma_{\rm min} < \gamma < \gamma_{\rm max}$,
we obtain
\begin{equation}
  n_\ast
  = \int_{\gamma_{\rm min}}^{\gamma_{\rm max}} 
        \frac{dn_\ast}{d\gamma} d\gamma
  = \frac{\gamma_{\rm min}^{1-p}-\gamma_{\rm max}^{1-p}}
           {p-1}
      n_0
\end{equation}
Provided that $\gamma_{\rm min} \ll \gamma_{\rm max}$ and $p>1$,
we obtain $n_0=(p-1)n_\ast$ (eq.~[\ref{eq:distr_nonth_1}]).
Assuming 
$\nu \gg (3/2) \gamma_{\rm min}^2 \nu_{\rm L}$ and
$\nu \ll (3/2) \gamma_{\rm max}^2 \nu_{\rm L}$,
we obtain the emission coefficient
\citep{Rybicki:1986:rpa},
\begin{eqnarray}
  j_\nu^{\rm (nt)}
  &=& 
    \frac{\sqrt{3}}{2} \frac{e^2}{c} \nu_{\rm L} n_0
    \frac{\displaystyle
          \Gamma \left( \frac{p}{4}+\frac{19}{12} \right)
          \Gamma \left( \frac{p}{4}-\frac{ 1}{12} \right)}{p+1}
  \nonumber \\
  &\times&
  \left( \frac13 \frac{\nu}{\nu_{\rm L}} \right)^{-(p-1)/2}
  \label{eq:emis_nt}
\end{eqnarray}
The absorption coefficient becomes
\citep{LeRoux:1961,Ginzburg:1965}
\begin{equation}
  \alpha_\nu^{\rm (nt)}
  = C(\alpha) \sin\chi \cdot n_0 r_0{}^2 \frac{\nu_0}{\nu_{\rm L}}
    \left( \frac{1}{\sin\chi} \frac{\nu_{\rm L}}{\nu} \right)^{2+p/2}.
  \label{eq:abso_nt}
\end{equation}

Once the distribution functions of thermal and nonthermal leptons,
as well as the magnetic-field distribution are specified,
we can compute the emission and absorption coefficients
(eqs.~\ref{eq:emis_th}, \ref{eq:abso_th}, \ref{eq:emis_nt} \&
      \ref{eq:abso_nt}) 
at each point in the jet,
and integrate the radiative transfer equation
along our line of sight to find the specific intensity.
We can then compute the distribution of 
the surface brightness on the celestial plane
to depict the expected VLBI map at each radio frequency, 
and compute the SED and the coreshift
as a function of frequency.

In the next section, 
we verify the {\tt R-JET} code
by comparing the resultant SEDs and coreshifts
with analytically inferred values.
For this purpose,
we assume that the jet material is composed of a pure pair plasma,
neglecting hadronic contribution.
In addition, we neglect the contribution 
of the local, tangled magnetic fields,
and consider only the global, ordered magnetic field
described in \S~\ref{sec:BZflux}.

\section{Code verification}
\label{sec:verification}
In this section,
we start with describing the jet evolution
\S~\ref{sec:evolution}.
Then we verify the {\tt R-JET} code 
by applying it to the thermal synchrotron process in \S~\ref{sec:CVS_thermal},
and the nonthermal synchrotron process in \S~\ref{sec:CVS_nonthermal}.
For the purpose of code verification, we assume a vanishing random 
magnetic field,
$B_{\ast,{\rm ran}}=0$, in \S\S~\ref{sec:CVS_thermal}--\ref{sec:CVS_nonthermal}.

\subsection{Example model of jet evolution}
\label{sec:evolution}
To constrain the lepton density in an axisymmetric jet,
we should specify the evolution of the bulk Lorentz factor $\Gamma$
and the magnetization parameter $\sigma$
as a function of $r$ along each magnetic flux tube.

In general, $\Gamma=\Gamma(r)$ should be specified source by source,
incorporating observational constraints.
However, in this particular section,
we assume that $\Gamma$ is constant along the jet
for the purpose of code verification.

As for the $\sigma$ evolution,
we follow an ideal MHD simulation for the M87 jet
\citep{Mertens:2016:AA} 
and assume a power-law dependence on $r$,
\begin{equation}
  \sigma(r)
  = \sigma_0 \left( \frac{r}{100 R_{\rm S}} \right)^{\sigma_{\rm p}}.
  \label{eq:sigma_evol}
\end{equation}
Outside the fast point, $\sigma$ continuously decreases
and attain $\sigma \ll 1$,
which means that the jet becomes kinetic-dominated asymptotically
\citep[e.g.,][]{Chiueh:1998:ApJ}.
We could mimic the $\sigma$ evolution of the M87 jet
\citet{Mertens:2016:AA}
by setting $\sigma_0 \approx 2.2$ and 
$\sigma_{\rm p} \approx -0.25$
in equation~(\ref{eq:sigma_evol}).
However, to apply the {\tt R-JET} code to other relativistic jets,
we leave the two parameters $\sigma_0$ and $\sigma_{\rm p}$ unconstrained
at this step,
and assume them appropriately problem by problem.

\subsection{Thermal synchrotron process}
\label{sec:CVS_thermal}
In this subsection, we assume a {\it thermal} plasma,
and adopt emission and absorption coefficients
described by equations~(\ref{eq:emis_th}) and (\ref{eq:abso_th}).
To compare with analytical solutions,
we assume that the BZ flux is constant for $A_\varphi$,
and apply $F_{\rm BZ,0}=T^r{}_t$ (eq.~[\ref{eq:Poynting_2}])
in equation~(\ref{eq:F_kin4}) for a test purpose.
We also assume that the bulk Lorentz factor $\Gamma$ is
constant for both $r$ and $A_\varphi$,
and adopt $\Gamma=2.0$ entirely in the jet.
The BH mass is assumed to be
$M= 6.4 \times 10^9 M_\odot$.

\begin{figure*}
\includegraphics[width=\textwidth, angle=0, scale=1.0]{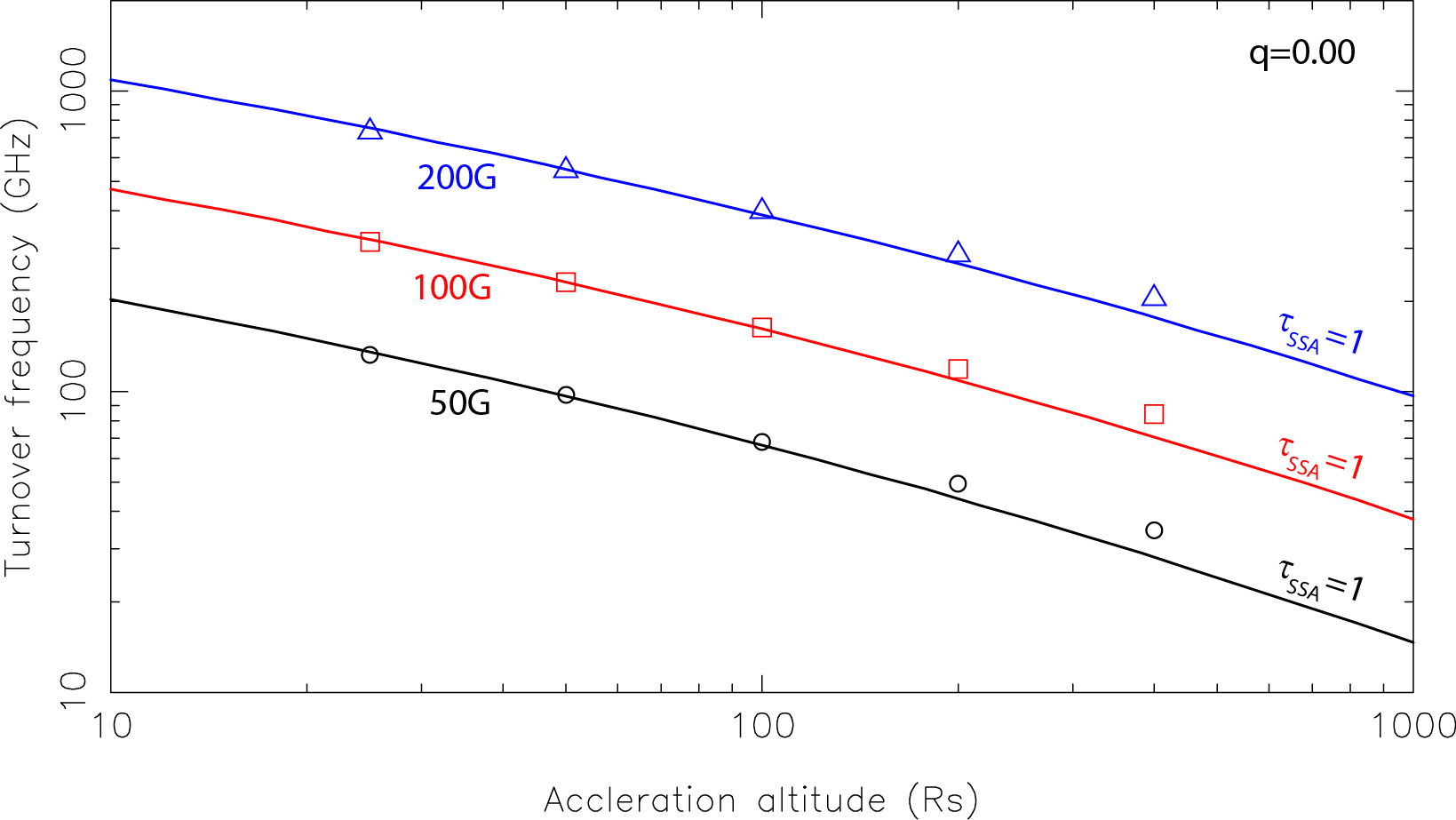}
\caption{
Turnover frequency of the synchrotron spectrum
of a thermal pair plasma as a function of the emission altitude. 
A conical jet, $q=0.00$, is assumed.
The lepton temperature is $\Theta_{\rm e}= 2.0$ in $m_{\rm e}c^2$ unit.
Analytical estimates (curves) deviate from the computed values 
with {\tt R-JET} (symbols),
because the analytical condition $\tau_{\rm SSA}=1$ estimates
the turnover frequency only approximately.
}
    \label{fig:turnover_1}
\end{figure*}

\begin{figure*}
\includegraphics[width=\textwidth, angle=0, scale=1.0]{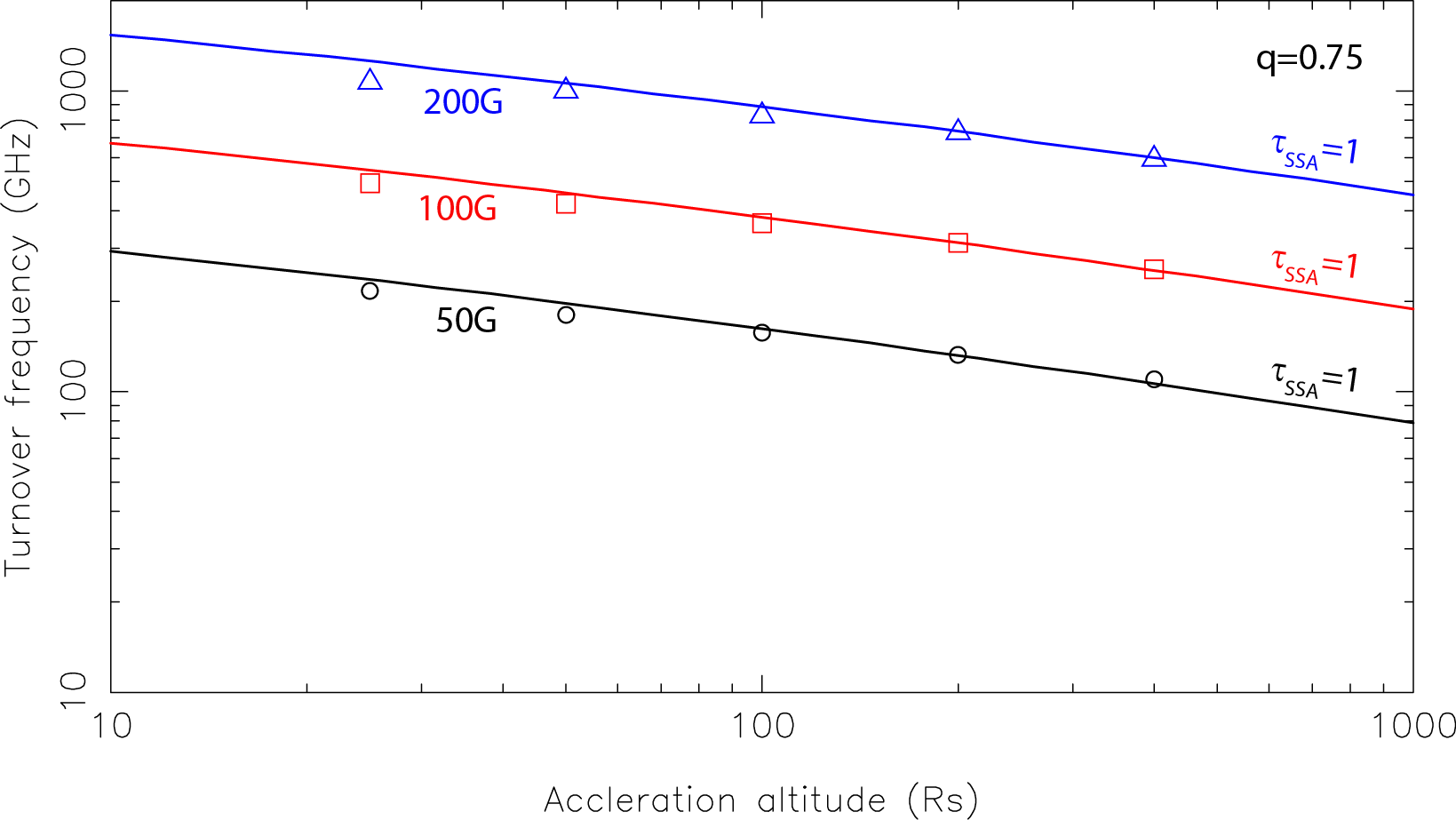}
\caption{
Similar figure as figure~\ref{fig:turnover_1}. 
A  quasi-parabolic jet, $q=0.75$, is assumed.
The lepton temperature is $\Theta_{\rm e}= 2.0$.
Analytical estimates (curves) deviate from the computed values (symbols)
due to a limitation of the condition $\tau_{\rm SSA}=1$.
}
    \label{fig:turnover_2}
\end{figure*}

In this particular subsection (\S~\ref{sec:CVS_thermal}),
to test the code,
we assume that thermal leptons exist only in a limited
spatial region.
Specifically, we consider five such cases in which
thermal leptons exist homogeneously 
within the radial range
$r_i < r < r_i + R_{\rm S}$,
where
$r_1= 25 R_{\rm S}$,
$r_2= 50 R_{\rm S}$,
$r_3= 100 R_{\rm S}$,
$r_4= 200 R_{\rm S}$, and 
$r_5= 400 R_{\rm S}$.
For example, in the first case, leptons
exist between the two concentric annuli
located at $r=25 R_{\rm S}$ and $r=26 R_{\rm S}$.
These five cases correspond to the symbols plotted at
$r/R_{\rm S}=25$, $50$, $100$, $200$, and $400$ 
in figures~\ref{fig:turnover_1}--\ref{fig:turnover_2}.
We adopt a semi-relativistic lepton temperature,
$\Theta_{\rm e}= 2.0$ in all the five cases.

We infer the turnover frequency in the SED
(in the unit of Jy) using the {\tt R-JET} code,
and compare the results with analytical estimates.
To analytically compute the turnover frequency,
$\nu_{\rm p}$,
we approximately infer it by setting $\tau_{\rm SSA}=1$,
where $\tau_{\rm SSA}= \alpha_\nu^{\rm (th)} \Delta$
denotes the absorption optical depth;
$\Delta= R_{\rm S} \sec\theta_{\rm v}$
denotes the path length of the emission region,
and $\theta_{\rm v}$ refers to the observer's viewing angle
with respect to the jet axis.
The actual value of $\tau_{\rm SSA}$ at which
the flux density peaks depends on the lepton energy distribution.
For instance, for {\it nonthermal} leptons,
the flux density peaks at smaller optical depths (than unity)
for a harder power-law energy distribution
\citep[e.g., \S~2 of][]{Hirotani:2005:ApJ}.
In the present case of {\it thermal} leptons,
to find $\nu_{\rm p}$ analytically,
we should integrate the radiative transfer equation along
each line of sight at each frequency,
then multiply the solid angle subtended by each segment in the celestial plane
to obtain the total flux density of the jet.
Then, we have to differentiate the result with respect to frequency
to obtain $\nu_{\rm p}$.
However, such an detailed analysis is out of the scope of the present paper,
and we approximately evaluate $\nu_{\rm p}$ simply by setting
$\tau_{\rm SSA}=1$,
knowing that there is an error in this analytical approach.

As the flow-line geometry, we consider two cases:
conical case ($q=0.00$) and quasi-parabolic case ($q=0.75$).
In figure~\ref{fig:turnover_1}, 
we show the results of $\nu_{\rm p}$ for $q=0.00$
as a function of the altitudes
at which the hot nonthermal leptons reside.
The black open circles, red open squares, and blue open triangles
represent the $\nu_{\rm p}$'s for 
$B_{\rm p,0}=50$~G, $100$~G, and $200$~G, respectively.
The solid curves denote their analytical values
estimated by $\tau_{\rm SSA}=1$.
It follows that the turnover frequency can be approximately
consistent with their analytical values.
Although the numerical values (open symbols) 
deviate from their analytical estimates
when nonthermal leptons reside $r > 200 R_{\rm S}$,
it is merely because $\tau_{\rm SSA}=1$ does not precisely give
the turnover frequency.
That is, the deviations are {\it not} due to the limitation of the code.

In the same manner, we adopt $q=0.75$ and compare the results
with their analytically inferred values.
Figure~\ref{fig:turnover_2} shows that the {\tt R-JET} code
gives consistent turnover frequencies (symbols)
with the analytical estimates (curves).
Again, the deviation from the analytical values (curves)
are due to the limitation of $\tau_{\rm SSA}=1$
as an analytical method to infer $\nu_{\rm p}$.

\subsection{Nonthermal synchrotron process}
\label{sec:CVS_nonthermal}
Next, let us check if the emission and absorption coefficients
are properly coded for {\it nonthermal} leptons.
To this end, 
we compare the computed coreshift
with the K$\ddot{\rm o}$nigl jet model
\citep{Blandford:1979:ApJ,1981ApJ...243..700K},
assuming a narrow conical jet.
In this model,
when the density and the magnetic-field strength evolves 
with the de-projected distance $r$ by a power law,
$n_{\ast,{\rm e}}^{\rm nt} \propto r^{-n}$, and
$B \propto r^{-m}$,
the coreshift depends on frequency $\nu$ as
\citep{Lobanov:1998:A&A}
$r_{\rm core}(\nu)= \Omega_{r\nu} \csc\theta_{\rm v} \nu^{-1/k_r}$,
where $\Omega_{r\nu}$ is defined by the normalization of
$B$ and $n_{\ast,{\rm e}}^{\rm tot}$
(e.g., by their values at $r=1$~pc)
as well as by the lepton energy distribution,
and does not depend on frequency $\nu$,
and analytically computable.
The quantity $k_r$ is defined by
\begin{equation}
  k_r \equiv \frac{(3-2\alpha) m + 2n - 2}{5-2\alpha},
  \label{eq:def_kr}
\end{equation}
where the optically thin power-law synchrotron index $\alpha$
is related to the lepton's power-law index $p$ by
$\alpha= (1-p)/2$.

\subsubsection{The case of toroidally-dominated magnetic field}
\label{sec:CVS_toroidal}
Let us begin with the case when
$\vert B_{\hat\varphi} \vert \gg B_{\rm p}$.
To examine a narrow conical jet, we set
$q=0$ and $\theta_{\rm BD}= 10^\circ$,
where $\theta_{\rm BD}$ denotes the colatitude
of the jet outer boundary (at the event horizon in general).
Because of the conical geometry (i.e., $q=0$),
the magnetic-field strength changes with $r$ as
$B_{\rm p} = (r/R_{\rm S})^{q-2} B_{\rm p,0}
 \propto r^{-2}$
on the poloidal plane.
On the other hand, its toroidal component behaves
$ \vert B_{\hat\varphi} \vert 
  = B_{\varphi,0} \sigma/(1+\sigma)
$,
where $B_{\varphi,0}$ denotes 
the toroidal component of the magnetic field at the horizon.
Let us assume $\sigma_{\rm p}= -1.0$ in \S~\ref{sec:CVS_toroidal}.
In this case, we obtain
$\vert B_{\hat\varphi} \vert \propto r^{-1}$.
Thus, $\vert B_{\hat\varphi} \vert$ 
decreases slowly compared to $B_{\rm p}$ with $r$.
Accordingly, if $\vert B_{\hat\varphi} \vert \gg B_{\rm p}$ 
holds at some inner point, say at $r= r_2$,
this relation holds at $r > r_2$.

Let us suppose $r_2= 20 R_{\rm S}$,
which will be sufficiently small compared to the jet entire scale.
The poloidal component becomes
$B_{\rm p}= 2.5 \times 10^{-3} B_{\rm p,0}$ 
at $r= 20 R_{\rm S}$.
On the other hand, the toroidal component becomes
$ \vert B_{\hat\varphi} \vert \approx
 B_{\rm p,0} \eta \sigma_0 (20/100)^{\sigma_{\rm p}}
 = 5 \sigma_0 \eta B_{\rm p,0}$
there.
Thus, as long as $\sigma_0 \gg 5 \times 10^{-4}$,
we obtain $\vert B_{\hat\varphi} \vert \gg B_{\rm p}$.

On these grounds, we  adopt $\sigma_0= 10^{-2}$ 
in \S~\ref{sec:CVS_toroidal}.
Note that we obtain $\sigma= 10^{-2}(20/100)^{-1}= 0.05$
at $r= 20 R_{\rm S}$, which guarantees
$\sigma \ll 1$ at $r > 20 R_{\rm S}$.
We thus assume that the jet's mass is loaded
at $r= 25 R_{\rm S}$, and all the photons are emitted
at $r > 25 R_{\rm S}$.
We examine the emission and absorption properties
of such a jet and investigate the brightness distribution
on the celestial plane.

\begin{figure*}
\includegraphics[width=\textwidth, angle=0, scale=1.0]{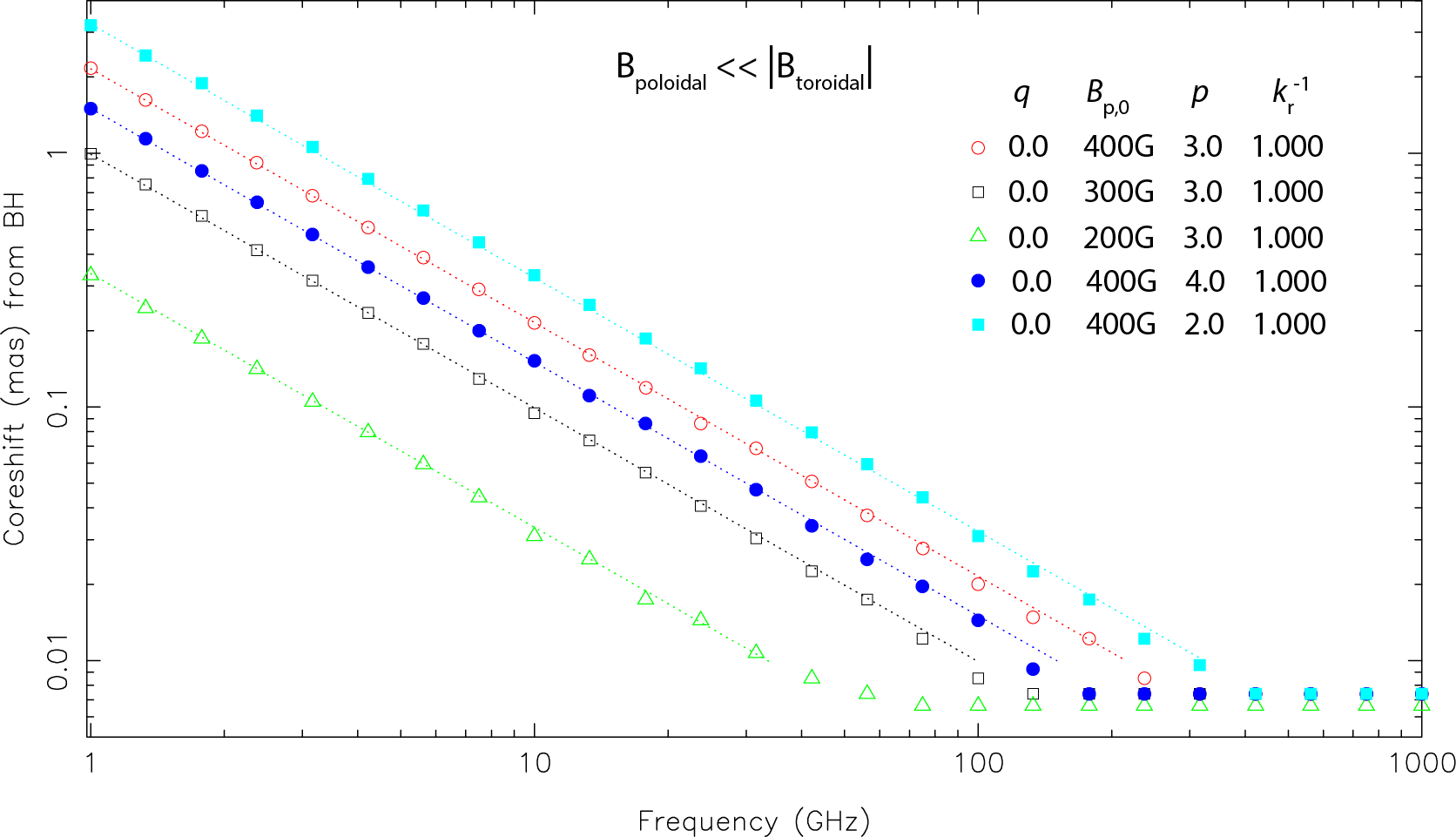}
\caption{
Coreshift, $r_{\rm core}$,
as a function of frequency for a narrow conical jet, 
$\theta_{\rm BD}=10^\circ$ and $q=0$,
when the toroidal component of the magnetic field
dominates the poloidal one.
The open and filled symbols show the numerical values
obtained for the set of $B_{\rm p,0}$ and $p$
as denoted in the top-right corner of the figure.
The thin dotted line show their analytical estimates
using the K$\ddot{\rm o}$nigl jet model.
Deviations appearing at $r_{\rm core} > 0.02$~mas (in the ordinate)
is due to a limitation in the accuracy of the {\tt R-JET} code.
}
   \label{fig:coreshift_3}
\end{figure*}

Figure~\ref{fig:coreshift_3}
shows the coreshifts computed with {\tt R-JET}
with open and filled symbols as a function of $\nu$.
The thin dotted lines show their analytical estimates
under the K$\ddot{\rm o}$nigl jet model.
We adopt five different combinations of $B_{\rm p,0}$ and $p$.
It is clear that the code gives consistent coreshifts
with their analytical estimates 
when the magnetic field is toroidally dominated.
Note that the computed coreshifts (i.e., open and close symbols) saturates at $0.00591$~mas in all the cases,
because the jet-launching point becomes optically thin
at higher frequencies.
Accordingly, the power-law dependence of the coreshift
is meaningful only at $r_{\rm core} > 0.02$~mas (in the ordinate).
Note that the evolution of the magnetic field and fluid density 
in this case (i.e., the case of $q=0$ and $\sigma_{\rm p}=-1.0$)
corresponds to what is expected for 
ideal, super-alfvenic MHD flows
along axisymmetric conical flow lines (without additional mass loading).

\subsubsection{The case of poloidally-dominated magnetic field}
\label{sec:CVS_poloidal}
Next, let us consider the case 
$B_{\rm p} \gg \vert B_\varphi \vert$.
To compare with the K$\ddot{\rm o}$nigl jet model,
we assume a narrow conical jet
and put $q=0.00$ and $\theta_{\rm BD}= 10^\circ$.
In this case, we find that the toroidal component of 
the magnetic field becomes
\begin{equation}
  \vert B_{\hat\varphi} \vert 
  \approx \eta B_{\rm p,0}  
          \frac{\sigma}{1+\sigma}
  \approx \eta B_{\rm p,0} \sigma,
\end{equation}
where $\sigma \ll 1$ is assumed again.
On the other hand, the poloidal component becomes
\begin{equation}
  B_{\rm p}= B_{\rm p,0} \left( \frac{r}{R_{\rm S}} \right)^{q-2}
\end{equation}
Thus, we obtain $B_{\rm p} \gg \vert B_\varphi \vert$
when the jet is strongly kinetic-dominated, $\sigma \ll 1$.
To ensure $B_{\rm p} \gg \vert B_{\hat\varphi} \vert$
in $r < 10^4 R_{\rm S}$,
we set $\sigma_0= 10^{-8}$ in \S~\ref{sec:CVS_poloidal}.

Under these assumptions, 
the magnetic field strength evolves with $r$ by
\begin{equation}
  B = \sqrt{ B_{\rm p}^2 + B_{\hat\varphi}^2}
    \approx B_{\rm p} \propto r^{q-2}.
  \label{eq:B_CVS}
\end{equation}
On the other hand, the proper lepton density evolves as
\begin{eqnarray}
  n_{\ast,{\rm e}}^{\rm tot}
  &=& \frac{1}{1+\sigma} \frac{B_{\rm p}(r,\theta)}{B_{\rm p,0}}
     \frac{F_{\rm BZ}(r,\theta)}{c \Gamma (\Gamma-1) U}
  \nonumber \\
  &\approx& \frac{B_{\rm p}(r,\theta)}{B_{\rm p,0}}
     \frac{F_{\rm BZ}(r,\theta)}{c \Gamma (\Gamma-1) U}
     \propto r^{q-2}.
  \label{eq:n_tot_CVS}
\end{eqnarray}
We thus find $n=m=q-2$ for this highly magnetically-dominated jet.
It is worth noting that this power-law dependence,
$n=m=q-2$, does not satisfy an energy equipartition 
between the radiating leptons and the global, ordered
magnetic field.
Nevertheless, since we assume an energy equipartition
between the leptons and the {\it random} magnetic field,
and since the random magnetic field
is suppressed in this particular section (\S~\ref{sec:verification}) for a test purpose,
this power-law dependence, $n=m=q-2$,
does not violate the assumptions made in the present work.

\begin{figure*}
\includegraphics[width=\textwidth, angle=0, scale=1.0]{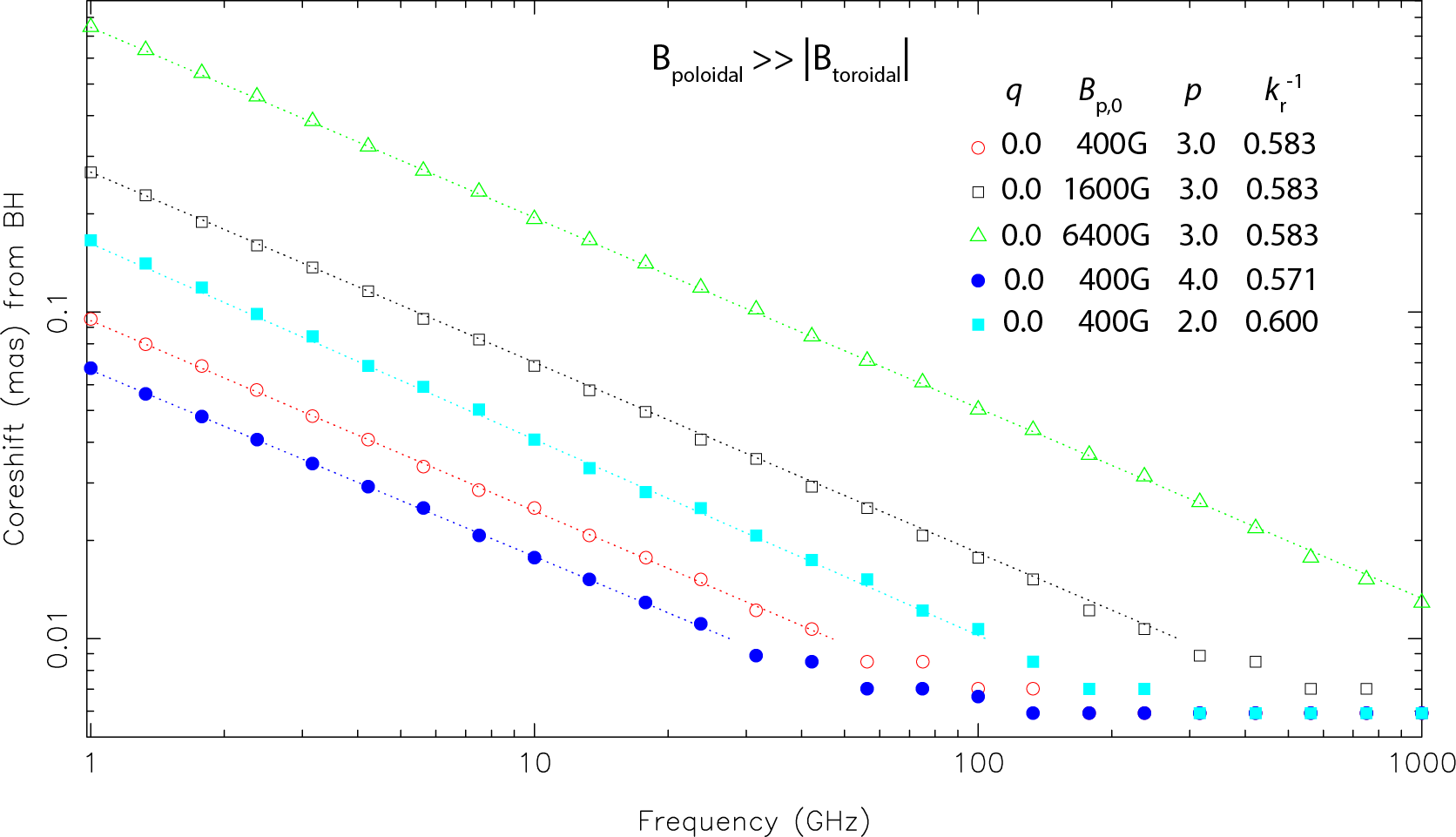}
\caption{
Coreshift, $r_{\rm core}$,
as a function of frequency for a conical jet, $q=0.0$,
when the poloidal component of the magnetic field
dominates the toroidal one.
The open and filled symbols show the numerical values
obtained for the set of $B_{\rm p,0}$ and $p$
as denoted in the top-right corner of the figure.
The thin dotted line show their analytical estimates
using the K$\ddot{\rm o}$nigl jet model.
Deviations appearing at $r_{\rm core} > 0.02$~mas 
is due to a limitation of the {\tt R-JET} code.
}
    \label{fig:coreshift_1}
\end{figure*}

We adopt the BH mass $M= 10^9 M_\odot$ in \S~\ref{sec:CVS_poloidal}.
Figure~\ref{fig:coreshift_1} shows
the resultant coreshift $r_{\rm core}$ 
as a function of frequency $\nu$ in the observer's frame.
The red open circles, black open squares, green open tirangles,
blue filled circles, and cyan filled squares
show the $r_{\rm core}$ obtained by the {\tt R-JET} code
for the parameter set tabulated in the top-right corner of the figure.
The power, $k_r{}^{-1}$, is computed from $q$ and $p$
by equation~(\ref{eq:def_kr}).
The power-law dependence of the coreshift on $r$
is meaningful only at $r_{\rm core} > 0.02$~mas 
by the same reason as explained in \S~\ref{sec:CVS_toroidal}.

Let us briefly discuss the values of $k_r$.
It follows from figure~\ref{fig:coreshift_1} that we obtain
$k_r \approx 1.7$ when the magnetic field is poloidally-dominated.
However, it is observationally suggested that $k_r$ becomes
typically between $0.6$ and $1.1$
\citep{hada11,Sokolovsky:2011:A&A,Kutkin:2014:MNRAS,Ricci:2022:A&A,
Nokhrina:2024:MNRAS}.
We consider that this discrepancy should be attributed 
to the assumption of a poloidally-dominated magnetic field 
when $\sigma \ll 1$, 
which contradicts with the MHD simulation results
that suggest poloidally-dominated global magnetic field 
for a Poynting-dominated jet ($\sigma \gg 1$) in the core region
\citep{Vlahakis:2004:ApJ, Lyubarsky:2009:ApJ, Komissarov:2009:MNRAS,
2009MNRAS.397.1486B,2023MNRAS.524.4012B}.
In this context, the cases considered in the foregoing subsection
(\S~\ref{sec:CVS_toroidal}) are more astrophysically reasonable 
compared to those in this subsection (\S~\ref{sec:CVS_poloidal}).
Let us restate that we consider a poloidally-dominated magnetic field 
with $\sigma \ll 1$ merely to check the code,
because the original K$\ddot{\rm o}$nigl jet model is also applicable
even to this case.

To sum up, the {\tt R-JET} code reproduces the analytical predictions 
both for thermal (\S~\ref{sec:CVS_thermal}) and nonthermal
(\S~\ref{sec:CVS_nonthermal}) leptons.

\section{Astrophysical implications}
\label{sec:implications}
In this section, we apply the {\tt R-JET} code to typical AGN jets
with realistic parameter choices.
Note that this section does not incorporate the specific assumptions
used for code verifications 
in \S\S~\ref{sec:CVS_thermal}--\ref{sec:CVS_nonthermal}.
Namely, we no longer assume that \,
(1) the BZ flux is constant for $A_\varphi$
(i.e., it is constant for $\theta$ at the horizon), \,
(2) the thermal leptons exist only in a limited spatial region, \,
(3) the flow lines are conical,
all of which were assumed in \S~\ref{sec:CVS_thermal}.
In the same way, we longer assume that \,
(1) the jet has a narrow conical shape 
(i.e., $q=0$ and $\theta_{\rm BD}=10^\circ$), \,
(2) $n_{\ast,{\rm e}}^{\rm tot} \propto r^{-n}$ 
(eq.~[\ref{eq:n_tot_CVS}]), \,
(3) $B \propto r^{-m}$ (eq.~[\ref{eq:B_CVS}]) \, 
(4) $\sigma_0= 10^{-8}$,
all of which were assumed in \S~\ref{sec:CVS_nonthermal}.
Instead, we return to the set up described in \S~\ref{sec:parabolic_jet}.

\subsection{Limb-brightened jets}
\label{sec:impl_limb}
AGNs commonly serve as the origin of highly collimated relativistic jets
with a parsec-scale structure characterized 
by a compact core and an extended jet.
The morphology of the extended jet generally shows a brightened central ridge
with an intensity profile concentrated along the jet axis.
However, 
limb-brightened structures have been found on smaller, sub-pc scales
from the jets of nearby radio galaxies, such as
Mrk~501 \citep{Giroletti:2004:ApJ,Piner:2009:ApJL,Koyama:2019:ApJ},
Mrk~421 \citep{Piner:2010:ApJ},
3C~84 \citep{Nagai:2014:ApJ, Giovannini:2018:NatAs, Savolainen:2023:A&A},
Cygnus~A \citep{Boccardi:2016:A&A},
M87 \citep{Walker:2018:ApJ,Lu:2023:Natur},
3C~264 \citep{Boccardi:2019:A&A},
Cen~A \citep{Janssen:2021:NatAs}, 
3C~273 \citep{Bruni:2021:A&A}, and 
NGC~315 \citep{Park:2024:ApJL}.

In the case of Cen~A,
its jet exhibits a central-ridge-brightened structure on larger, pc scales
\citep{Horiuchi:2006:PASJ,Ojha:2010:A&A,Muller:2014:A&A}.
However, a recent observation of Cen~A with 
The Event Horizon Telescope (EHT) at 1.3~mm
uncovered a jet structure featuring limb-brightening on sub-pc scales
\citep{Janssen:2021:NatAs}.
A likely explanation for this contrast is that the jet is
inherently limb-brightened,
and that this feature has remained undetectable due to the limited angular 
resolution of earlier VLBI observations at lower frequencies.
What is more,
limb-brightened jets are so far detected
from nearby AGNs or from the AGNs harboring substantial BH masses
\citep[e.g., see \S~1 of ][]{Park:2024:ApJL}.
It may be, therefore, reasonable to suppose that
AGN jets are generally limb-brightened on sub-pc and pc scales,
and that most of them have been observed to display central-ridge-brightened
structures, largely due to the restricted angular resolution
of earlier VLBI observations.

To investigate how such brightness patters are produced in jets,
theoretical models have been proposed using force-free or MHD models.
For instance, 
integrating the radiative transfer equation in 
over-pressured super-fast-magnetosonic jets,
\citet{Fuentes:2018:ApJ} 
investigated the images of jets with transverse structure and knots 
with a large variety of intensities and separations.
Then, assuming a ring-like distribution of emitting electrons
at the jet base, 
\citet{Takahashi:2018:ApJ} investigated the formation
of limb-brightened jets in M87, Mrk~501, Syg~A, and 3C~84,
and showed that symmetric intensity profiles require
a fast spin of the BH.
Subsequently, \citet{Ogihara:2019:ApJ} 
examined a steady axisymmetric force-free jet model, 
and found that the fluid's drift velocity produces the 
central ridge emission due to the relativistic beaming effect,
and that the strong magnetic field and high plasma density 
near the edge results in a brightened limb.
\citet{Kramer:2021:A&A} applied 
a polarized radiative transfer and ray-tracing code to 
synchrotron-emitting jets, and found that the jet becomes limb-brightened
when the magnetic field is toroidally dominated, and that 
the jet becomes spine-brightened when the field is poloidally dominated.
More recently, examining the transverse structure of MHD jet models,
and computing radiative transfer of synchrotron emission and absorption,
\citet{Frolova:2023:MNRAS} showed that triple-peaked transverse profiles
constrain the fraction of emitting leptons in a jet.

Following these pioneering works, and utilizing 
MHD simulations in the literature,
H24 applied the initial version of the {\tt R-JET} code to the M87 jet.
Assuming that the jet is composed of a pure pair plasma
with a hybrid thermal and nonthermal energy distribution,
H24 demonstrated 
that a limb-brightened structure will be naturally formed
within the initial pc scales
by virtue of the angle-dependent energy extraction from 
the rapidly rotating supermassive BH in the center of M87 galaxy.

In the present paper, we apply the revised version
of the {\it R-JET} code, which allows more flexible settings
of jet structures (e.g., global and local magnetic field distributions) 
and parameters (e.g., the lepton fraction).
In this subsection (\S~\ref{sec:impl_limb}), 
we consider a pure pair plasma in the same way as H24,
but assume that all the pairs are nonthermal.
In addition, to consider a simple situation, 
we consider a quasi-parabolic jet
whose power-law index of $A_\varphi$ is
spatially constant at $q=0.75$,
whereas $q$ is assumed to evolve with $r$ in H24.
Since the leptons are nonthermal, 
we do not define the temperature, $\Theta_{\rm e}$
in this subsection (\ref{sec:impl_limb}).
We define the jet boundary as the magnetic flux surface
threading the horizon on the equatorial plane.
Accordingly, in the polar coordinates ($r$,$\theta$),
the jet boundary is defined by
$ (r/R_{\rm S})^q (1-\cos\theta)= 1$,
where we neglect GR effects in the {\tt R-JET} code.
We then apply the {\tt R-JET} code to the quasi-parabolic jet
and compute the surface brightness distribution on the celestial plane.
We assume that the BH has a mass $M=10^9 M_\odot$ and a spin $a=0.9M$,
and is located at the angular diameter distance $d_{\rm A}= 10$~Mpc.
For presentation purpose, we select a viewing angle
$\theta_{\rm v}=30^\circ$ with respect to the jet axis.
At $10$~Mpc, one Schwarzschild radius, $R_{\rm s}$, corresponds to
the angular diameter $2.0 M_9$~$\mu$as,
where $M_9$ refers to the BH mass measured in $10^9 M_\odot$ unit.

\begin{figure*}
\includegraphics[width=\columnwidth, angle=0, scale=1.0]{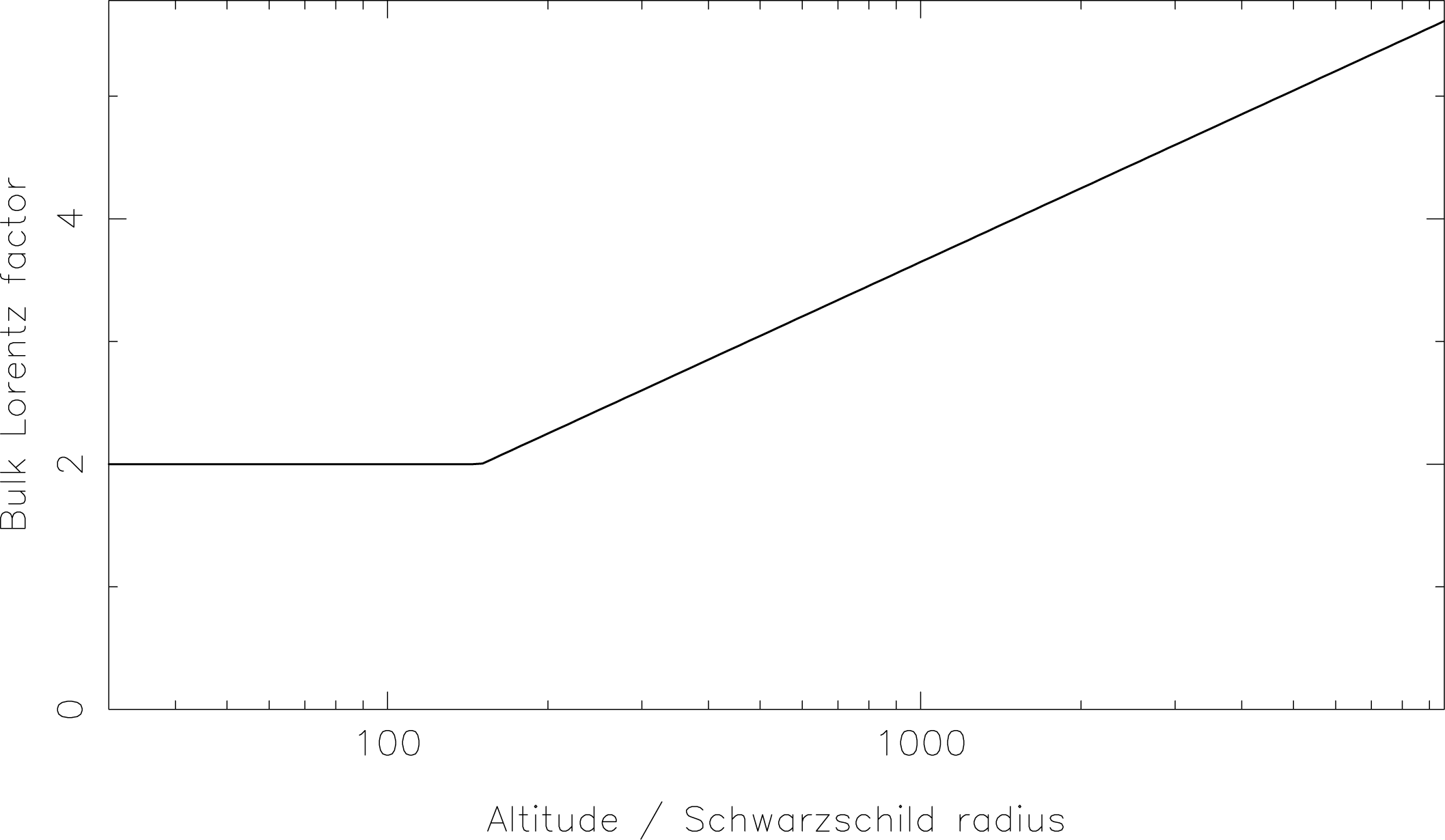}
\includegraphics[width=\columnwidth, angle=0, scale=1.0]{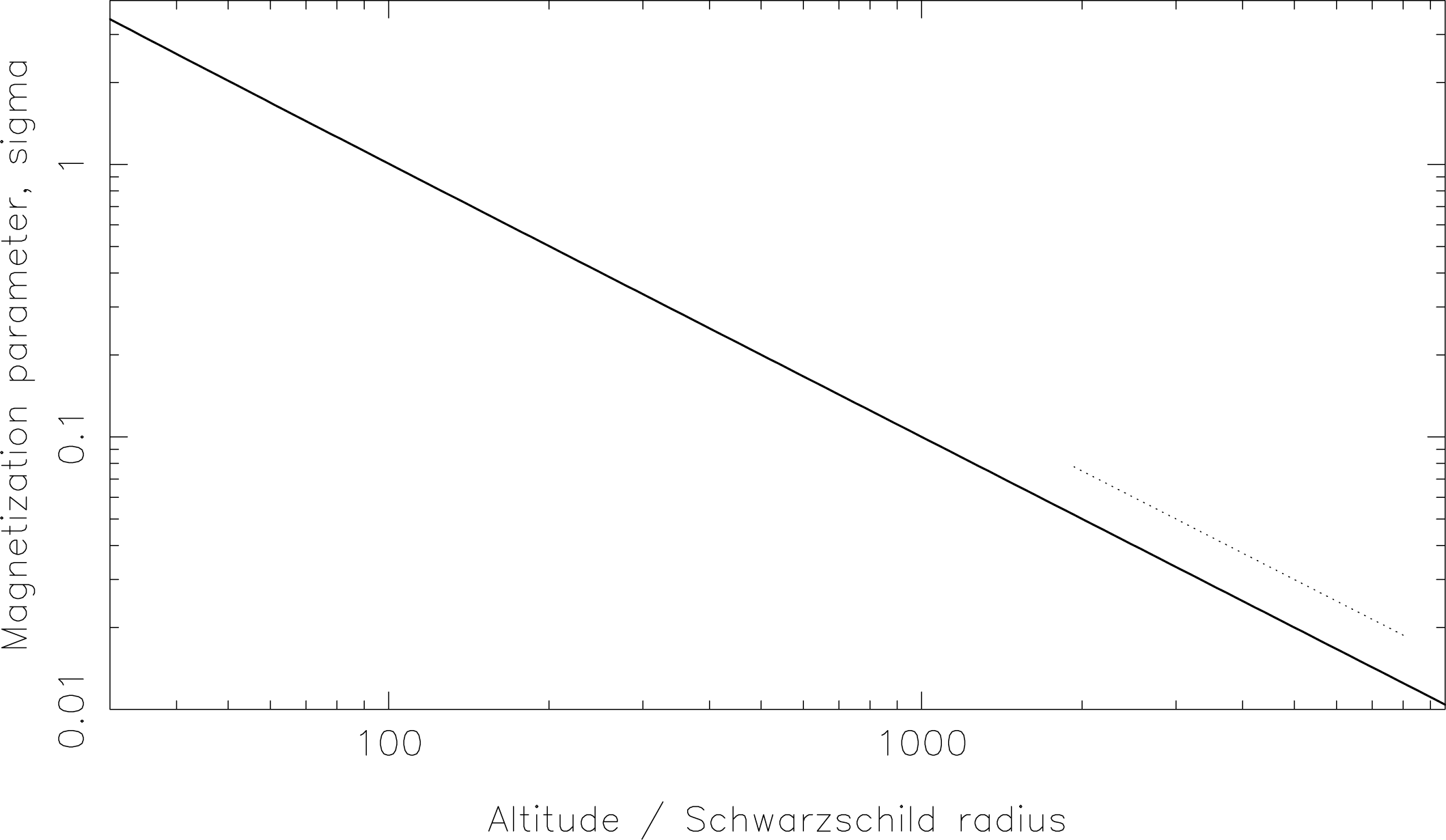}
\caption{
Radial distribution of the bulk Lorentz factor (left panel) and 
the magnetization parameter (right panel).
Both quantities are assumed to take the same value
along different magnetic flux tubes at each
distance $r$ from the BH.
}
    \label{fig:Gamma_sigma}
\end{figure*}

Let us begin with describing the magnetic field distribution
in the jet.
The left and right panels of figure~\ref{fig:Gamma_sigma}
show the radial variation of the bulk Lorentz factor $\Gamma$
and the magnetization parameter $\sigma$, respectively.
Both quantities are input parameters,
and assumed to have no dependence on $A_\varphi$
(i.e., depend on $r$ alone in the poloidal plane).

\begin{figure*}
\includegraphics[width=\columnwidth, angle=0, scale=1.0]{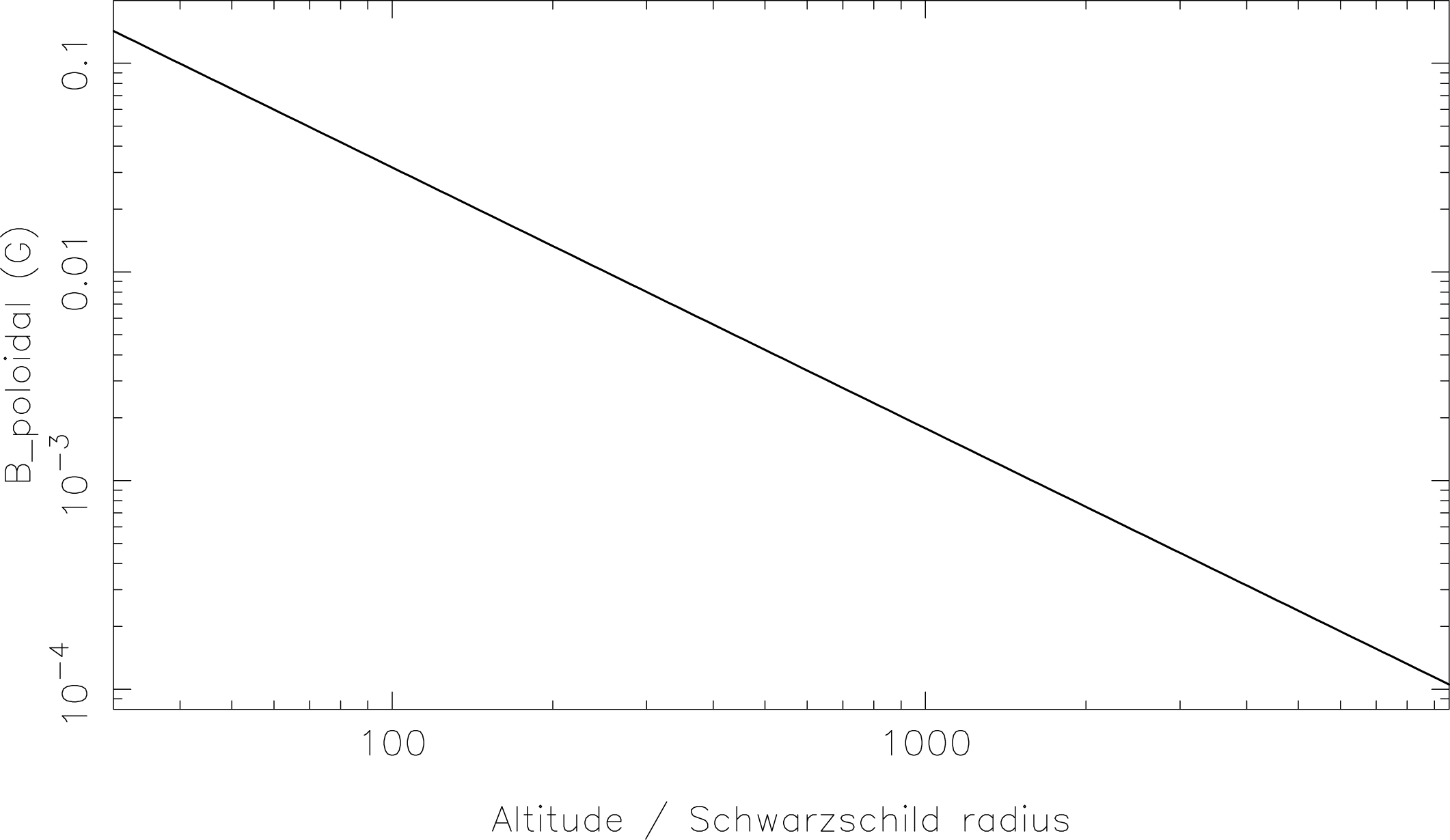}
\includegraphics[width=\columnwidth, angle=0, scale=1.0]{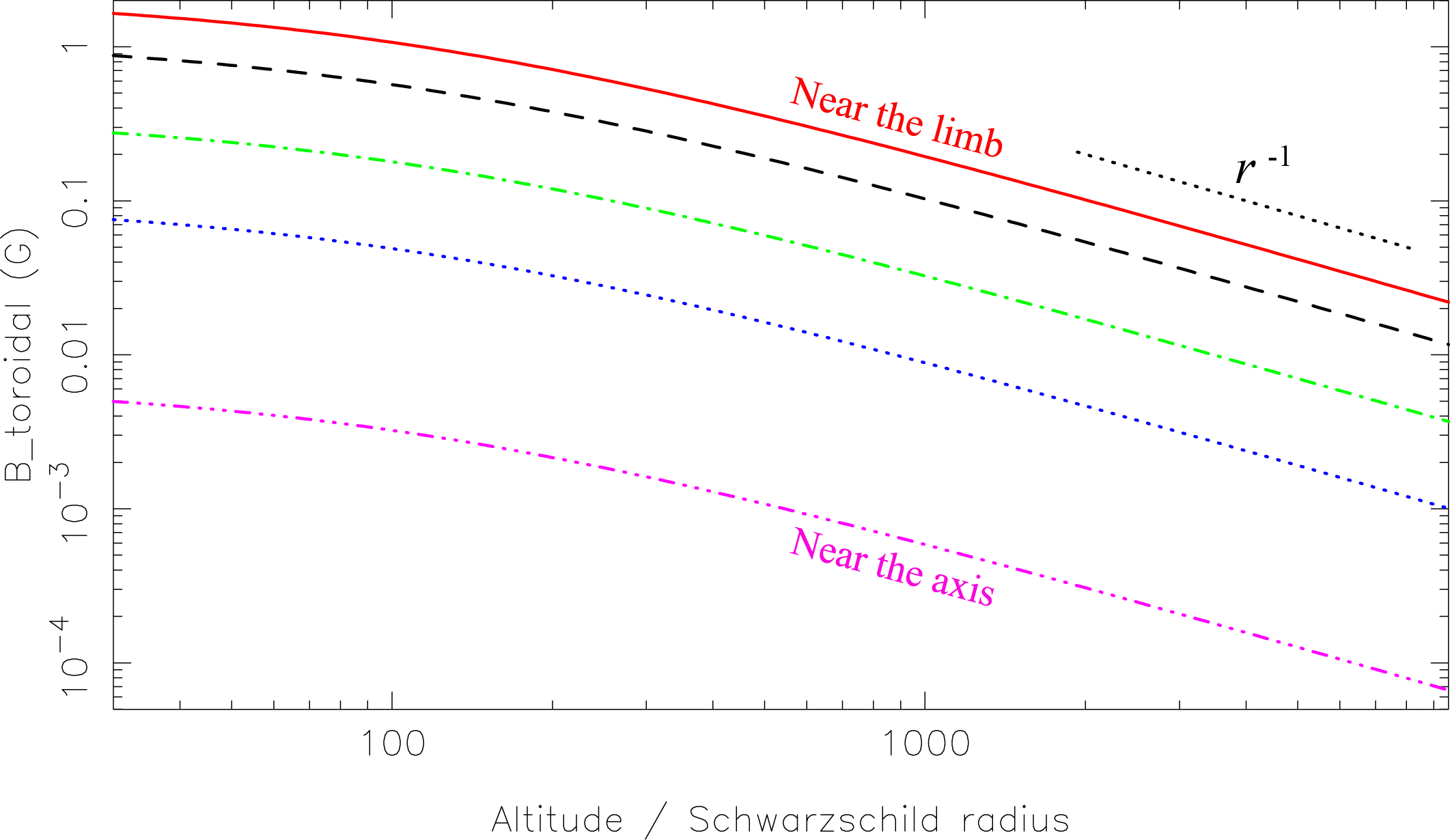}
\caption{
Radial distribution of the poloidal (left panel) and 
the toroidal (right panel) components of the ordered magnetic field.
For the poloidal component, meridional dependence is 
negligible, because $\theta \ll 1$ at $r \gg R_{\rm S}$ 
in equation~(\ref{eq:Bp}).
For the toroidal component, meridional dependence
is shown by the five curves:
the red solid, black dashed, green dash-dotted,
blue dotted, and purple dash-dot-dot-dotted curves
show the physical component $B_{\rm \hat\varphi}$ 
in the AGN-rest frame
at $A_{\varphi}= 0.96875$, $0.50000$, $0.25000$, $0.12500$,
and $0.03125$.
}
    \label{fig:Bpol_Btor}
\end{figure*}

The left panel of figure~\ref{fig:Bpol_Btor} shows
$B_{\rm p}$,
the strength of the large-scale, ordered magnetic field
in the poloidal plane.
At $r \gg R_{\rm S}$, $\theta \ll 1$ results in
a negligible dependence of $B_{\rm p}$ on $\theta$
(or equivalently, on $A_\varphi$).
We thus present the value of $B_{\rm p}$ along the 
jet axis, where $\theta= A_\varphi= 0$ holds.
The right panel shows the variation of 
the toroidal magnetic field $B_{\hat\varphi}$,
which represents the physical magnetic-field strength in gauss.
The purple dash-dot-dot-dotted, blue dotted,
green dash-dotted, black dashed, and red solid
curves show the values of $B_{\hat\varphi}$
along the magnetic flux tube with
$A_\varphi / A_{\varphi,{\rm max}}= 0.031$, 
$0.125$, $0.250$, $0.500$, and $0.968$, respectively,
where the magnetic-flux surface with 
$A_\varphi= A_{\varphi,{\rm max}}$ defines the jet boundary.
As expected, the magnetic field is wound more and more
strongly towards the limb, where the total energy flux
(i.e., the sum of the electromagnetic and kinetic fluxes)
peaks across the jet.
Both $B_{\rm p}$ and $B_{\hat\varphi}$ are input parameters
in the sense that their spatial distribution is analytically
constrained once $q$ and $\sigma=\sigma(r)$ are specified.

\begin{figure*}
\includegraphics[width=\columnwidth, angle=0, scale=1.0]{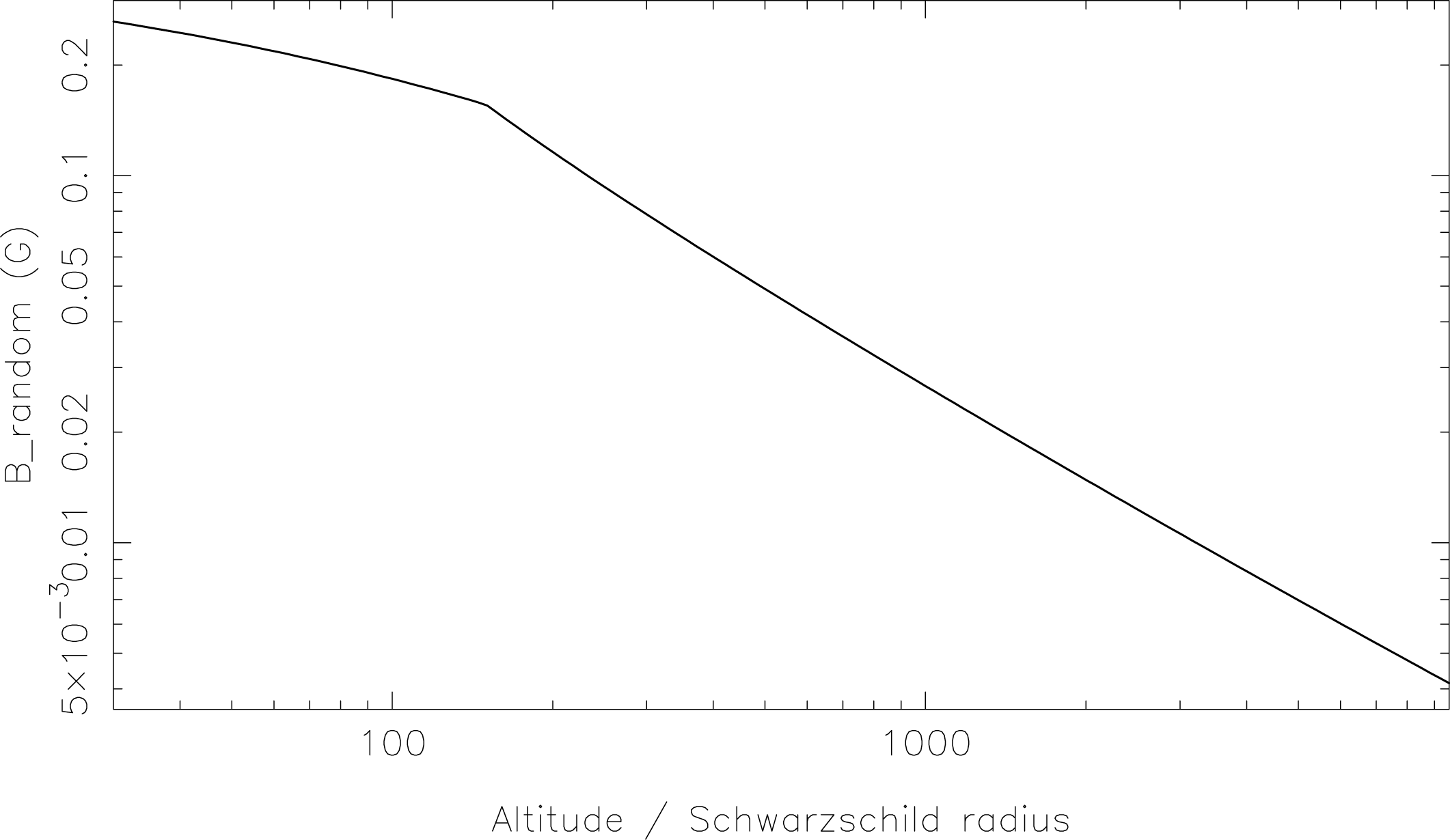}
\includegraphics[width=\columnwidth, angle=0, scale=1.0]{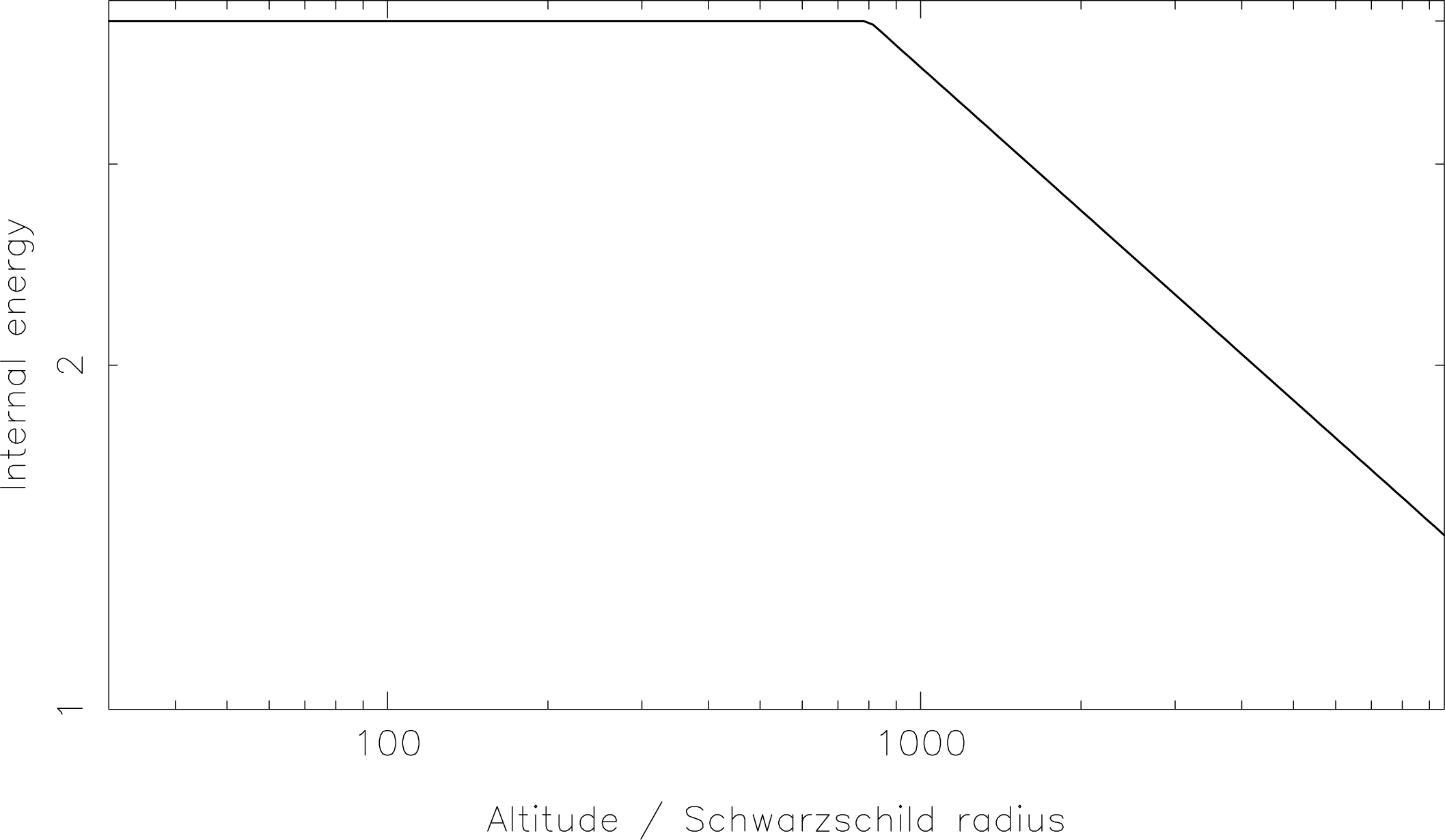}
\caption{
Radial distribution of the random magnetic-field strength (left panel) 
and the random energy of pairs (right panel).
Both quantities are assumed to have no dependence
on the magnetic flux function, $A_\varphi$.
The random magnetic field strength kinks at 
$r=150 R_{\rm S}$, because the bulk Lorentz factor
(fig.~\ref{fig:Gamma_sigma}) is assumed to increase
from that altitude.
The random energy kinks at $r=800 R_{\rm S}$,
because the nonthermal pairs are assumed to cease
acceleration at that altitude,
and adiabatically cooled down at $r > 800 R_{\rm S}$.
}
    \label{fig:Bran_Eran}
\end{figure*}

Figure~\ref{fig:Bran_Eran} shows the radial dependence
of $B_{\ast,{\rm ran}}$ and $U/m_{\rm e}c^2$.
By definition, neither quantities depend on $A_\varphi$.
The random magnetic field has a kink at $r=150 R_{\rm S}$
due to the Lorentz factor distribution
(left panel of fig.~\ref{fig:Gamma_sigma}).
The random-motion energy of nonthermal particles
(in the present case, electrons and positrons)
decreases at $r>800 R_{\rm S}$, 
because we assume that pairs are continuously accelerated
at $r<800 R_{\rm S}$ and begin to be cooled down
mainly via adiabatic expansion and partly via 
synchrotron cooling at $r>800 R_{\rm S}$.
Precisely speaking, high-energy tail leptons
preferentially lose energy via synchrotron cooling;
however, due to the steep power-law energy distribution
(e.g., $p=3.0$ in the present case),
leptons lose internal energies mainly via adiabatic expansion
as a whole.

\begin{figure*}
\includegraphics[width=\textwidth, angle=0, scale=1.0]{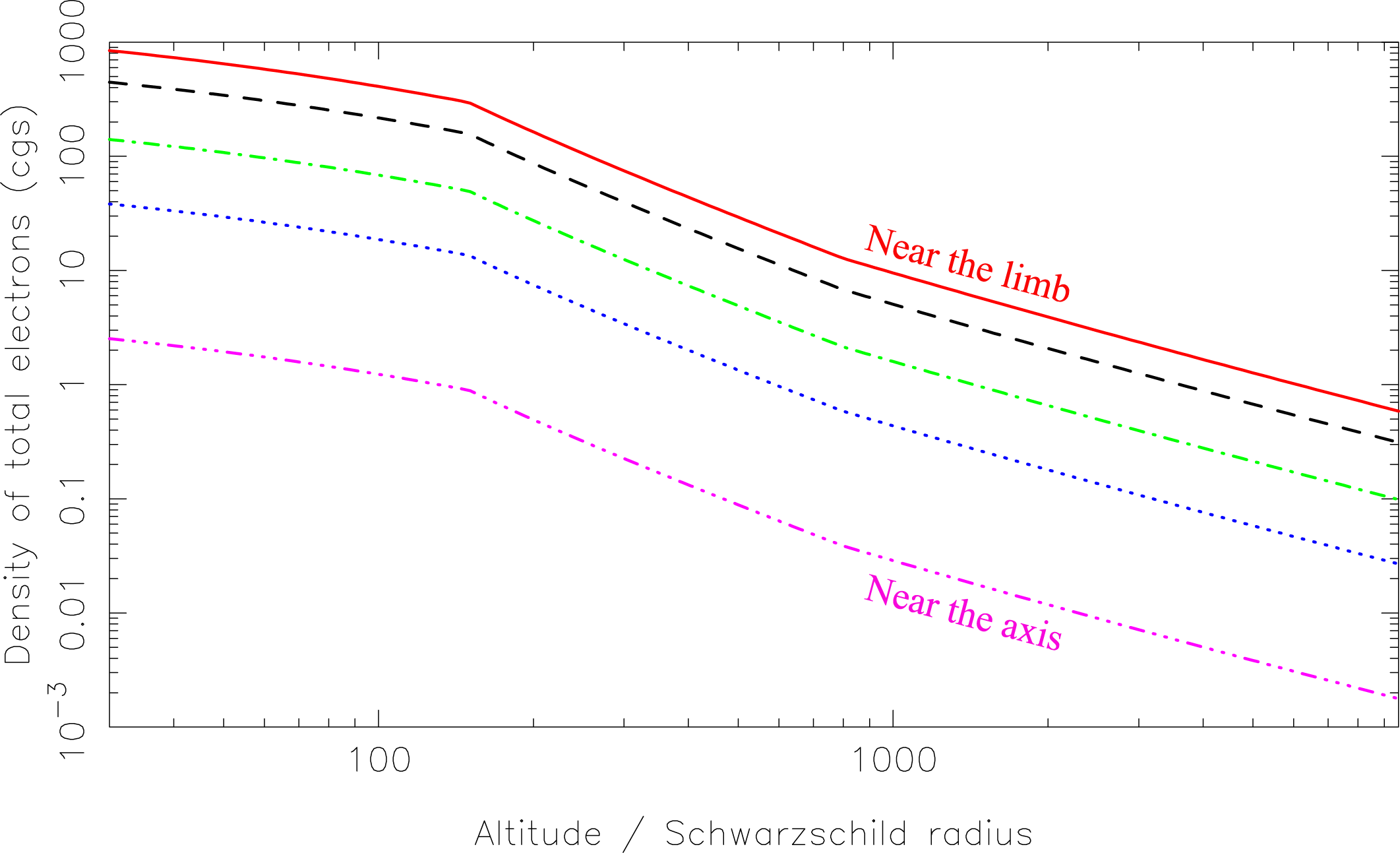}
\caption{
Radial distribution of the electron number density
in $\mbox{cm}^{-3}$.
The red solid, black dashed, green dash-dotted,
blue dotted, and purple dash-dot-dot-dotted curves
show the $n_{\ast,\rm e}^{\rm tot}$ 
at $A_{\varphi}= 0.96875$, $0.50000$, $0.25000$, $0.12500$,
and $0.03125$.
}
    \label{fig:Ne_tot}
\end{figure*}

Figure~\ref{fig:Ne_tot} shows the variation 
of the total lepton number density, 
$n_{\ast,{\rm e}}^{\rm tot}$, 
in the jet co-moving frame.
The curves corresponds to the same magnetic flux surfaces
as in the right panel of figure~\ref{fig:Bpol_Btor}.
Because we assume that the initial Poynting flux is
efficiently converted into the kinetic flux,
the particle number density rapidly increases toward 
the jet limb,
leading to a large emissivity there.

\begin{figure*}
\includegraphics[width=\columnwidth, angle=0, scale=0.7]{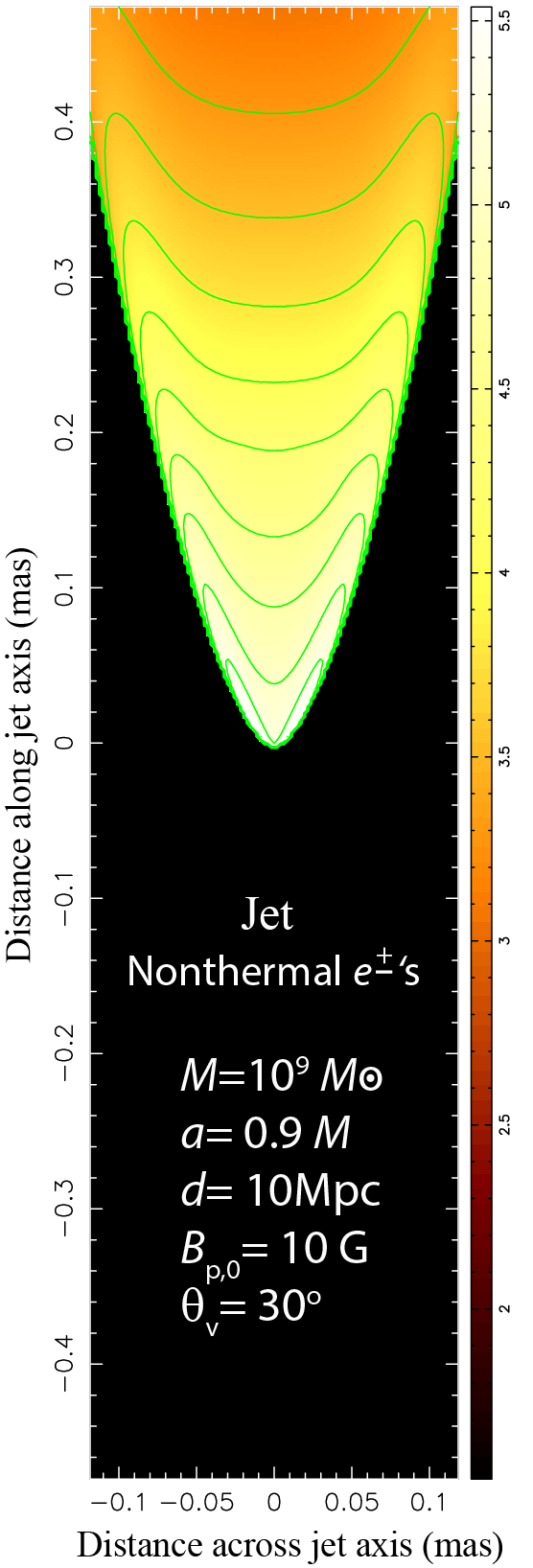}
\includegraphics[width=\columnwidth, angle=0, scale=0.7]{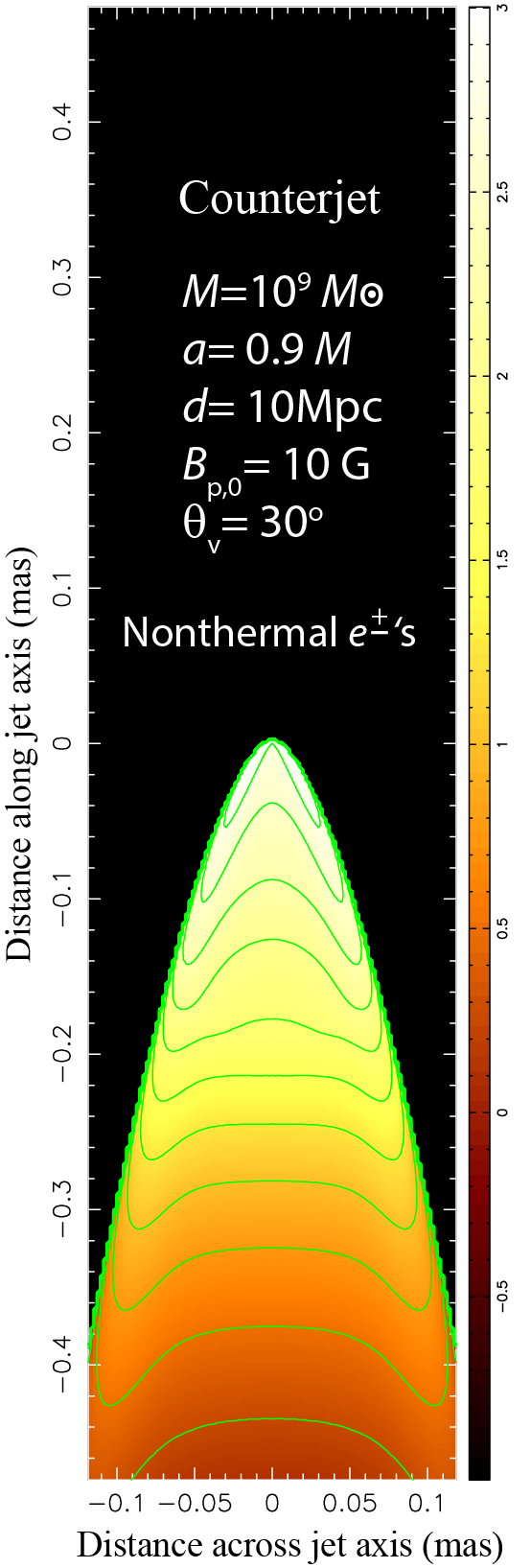}
\caption{
Surface brightness distribution of the jet.
The origin (0,0) shows the line of sight toward the BH
whose mass and spin are $M=10^9 M_\odot$ and $a=0.9M$.
The magnetic-field strength is $B_{\rm p,0}= 10$~G
at the jet base.
The distance is assumed to be $10$~Mpc,
and the observer's viewing angle is $\theta_{\rm v}= 30^\circ$.
The left panel corresponds to the approaching jet,
while the right one does the receding jet.
It is assumed that the plasma is composed of a pure pair plasma
($f_{\rm p}= 1.0$), 
and that their energy distribution is nonthermal
with a power-law index $p=3.0$.
The photon frequency is $230$~GHz in the observer's frame.
The color is coded logarithmically in $\mu \mbox{Jy mas}^{-2}$ unit;
e.g., $4.0$ in the color bar (on the right) corresponds to
$10^4 \mu \mbox{Jy mas}^{-2}= 10^{-2} \mbox{Jy mas}^{-2}$.
The peak brightness is $344 \mbox{mJy mas}^{-2}$ and
$1.00 \mbox{mJy mas}^{-2}$ for the left and right panel, respectively.
}
    \label{fig:map_jet}
\end{figure*}

We next show that this high plasma density
and the strong $B_{\hat\varphi}$ at the jet limb
results in a limb-brightened structure.
Adopting the photon frequency $\nu= 230$~GHz in the observer's frame,
we obtain the expected VLBI map of this jet as
presented in figure~\ref{fig:map_jet}.
The left and right panels show the surface brightness of 
the approaching and receding jets, respectively.
It follows that the jet appears limb-brightened as a result of 
the angle-dependent energy extraction from a rotating BH.
In this subsection (\S~\ref{sec:impl_limb}),
we assume that the leptons are continuously accelerated
and the power-law index and the upper cutoff Lorentz factor
is spatially constant;
accordingly, only the normalization (i.e., the density)
of leptons decreases due to the expansion of the fluid element
toward the jet downstream.
As a result, the brightness slowly decreases 
with distance from the BH.
For a wide range of parameters, such as for 
various BH masses and spins,
various viewing angles,
various fractions of nonthermal pairs, as well as
various compositions of plasmas 
(i.e., whether the jet is pair-plasma dominated or normal-plasma dominated), 
we find that limb-brightened jets are formed universally.
Thus, we conclude that the limb-brightening is a general feature
of BH jets,
as long as they are energized by the BH's rotational energy
via the BZ process.

\subsection{Formation of a ring-like structure}
\label{sec:impl_ring}
At 230~GHz, EHT revealed
that the center of the M87 galaxy (namely, M87*)
exhibits a ring-like structure of a brightened region
with a diameter
approximately $5$ times Schwarzschild radii ($R_{\rm S}$)
\citep{EHT:2019:ApJL:1,EHT:2019:ApJL:6}.
The ring-like structure was confirmed at a lower frequency,
86~GHz, by \citet{Lu:2023:Natur},
with a greater diameter $\approx 8.4 R_{\rm S}$.
EHT also found a ring-like structure 
with a diameter $51.8 \mu$as from the radio core of Sgr A*
\citep{EHT:2022:ApJL:1,EHT:2022:ApJL:4}

Such ring-like structures are considered
as the signature of the existence of the event horizon,
because the gravitationally lensed photon orbits
are suggested to result in such a brightness distribution
\citep{Jaroszynski:1997:A&A,Falcke:2000:ApJL}.
Nevertheless, the ring-like structure of the M87*
observed at 86~GHz shows approximately $50$~\% greater diameter
than the EHT's photon ring at 230~GHz,
and may be connected to its limb-brightened jet
\citep[see the arguments in][]{Lu:2023:Natur}.

\begin{figure*}
\includegraphics[width=\columnwidth, angle=0, scale=0.7]{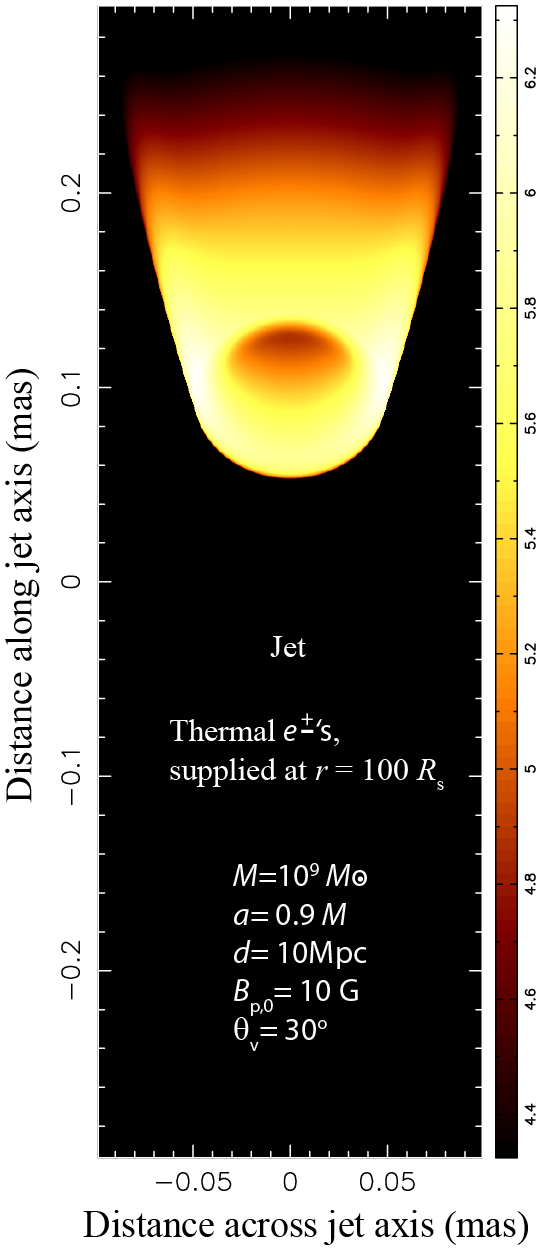}
\includegraphics[width=\columnwidth, angle=0, scale=0.7]{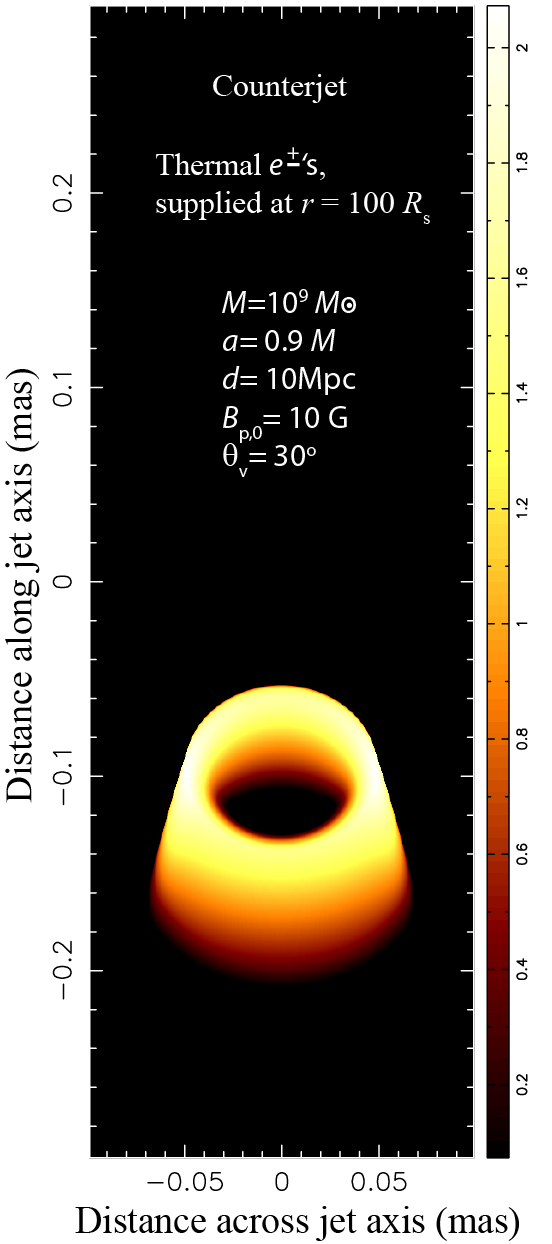}
\caption{
Surface brightness distribution of the jet base.
The color is coded logarithmically in $\mu \mbox{Jy mas}^{-2}$ unit.
Thermal leptons are supplied at $r=100 R_{\rm S}$ 
with relativistic temperature, $\Theta_{\rm e}= 10.0$.
The left panel corresponds to the approaching jet,
while the right one does the receding jet.
The photon frequency is $230$~GHz in the observer's frame.
The peak brightness is $2113 \mbox{mJy mas}^{-2}$ and
$0.1182 \mbox{mJy mas}^{-2}$ for the left and right panel, respectively.
The ring-like structure appears because we observe a limb-brightened
jet nearly face-on.
}
    \label{fig:map_ring}
\end{figure*}

In this subsection (\S~\ref{sec:impl_ring}), 
we thus consider if such a ring-like structure is formed 
by the synchrotron emission from a limb-brightened jet.
Applying the {\tt R-JET} code to the same parameter set
as in \S~\ref{sec:impl_limb},
and assuming that the jet plasmas are purely thermal
and supplied with a relativistic temperature 
$\Theta_{\rm e}=10.0$ at $r= 100 R_{\rm S}$,
we obtain the expected VLBI map as presented in
figure~\ref{fig:map_ring}
at $\nu=230$~GHz in the observer's frame 
(i.e., in our frame of reference).
Because of the angle-dependent energy extraction from the BH
(fig.~1 of H24),
the synchrotron photons emitted from the limb-brightened jet base 
(\S~\ref{sec:impl_limb}) exhibit a ring-like
structure of surface brightness in the celestial plane,
because we observe the limb-brightened jet nearly face-on.
In the present case, 
the diameter of the ring-like structure is approximately $0.1$~mas,
which is about $5 R_{\rm S}$,
by virtue of a strong collimation of the jet with $q=0.75$.
If we fix the height of plasma supply
(at $100 R_{\rm S}$ in the present case),
the ring diameter increases (or decreases)
with decreasing (or increasing) $q$,
because the flow-line geometry approaches
a less collimated, conical shape
(or a more collimated, parabolic shape).
The brightness rapidly decreases in the jet downstream
with the distance from the BH,
because the relativistic leptons lose energy 
by adiabatic expansion.
Unlike the photon ring observed with EHT,
this ring-like structure does not directly show 
the existence of the event horizon,
and may correspond to the ring-like structure found
by \citet{Lu:2023:Natur}.
Nevertheless, this ring-like structure may indicate the fact
that jets are energized by the BZ process,
which extracts the BH's rotational energy
in an angle-dependent way
(e.g., fig.~1 of H24).

\subsection{Impact of hadronic contribution}
\label{sec:impl_content}
The particles injected into a jet may have their origin
not only in a relativistic {\it pair} plasma,
but also in a semi-relativistic {\it normal} plasma.
The former, pair plasma could be supplied 
by a pair-production cascade in a BH magnetosphere
\citep{bes92,Hirotani:1998:ApJ,nero07,levi11,Broderick:2015:ApJ,
hiro16a,Kisaka:2020:ApJ}.
On the other hand, the latter, normal plasma could be supplied 
by an advection from a RIAF,
and consist of protons and electrons
as long as helium and heavier elements are ignored.
In this section, we thus investigate 
the impact of a hadronic contamination in a jet,
assuming a pure-hydrogen plasma as the normal plasma.
We can incorporate the normal plasmas in the code
by setting $f_{\rm p}<1$ in equation~(\ref{eq:def_U}).

We start with considering the value of $f_{\rm p}$ below which 
a hadron contamination contributes significantly.
For the present case of $p \approx 3$,
the averaged Lorentz factor of nonthermal pairs becomes
$\langle \gamma \rangle \approx 2 \gamma_{\rm min}$,
provided that $\gamma_{\rm max} \gg \gamma_{\rm min}$.
Therefore, the proton's inertia affects the lepton density through
equations~(\ref{eq:F_kin}) and (\ref{eq:def_U})
if $f_{\rm p}$ is less than or nearly equals to
$1 - (2\langle\gamma\rangle)/1836 \approx 0.998$
for $\gamma_{\rm min} \approx 1$.
On the other hand, if $p \approx 2$,
the harder lepton's energy distribution results in
a greater lepton mass, $\langle\gamma\rangle \approx 10$,
in the jet co-moving frame.
Owing to this heavy lepton mass,
proton mass contributes to reduce the pair density
significantly when $f_{\rm p} < 0.99$ if $p \approx 2$.

On these grounds, we choose
$f_{\rm p}= 1.00$, $0.99$, $0.90$, $0.50$, and $0.10$,
as representative values, and compare the resultant spectra.
We adopt the same parameter set as \S~\ref{sec:impl_ring};
namely, $M=10^9 M_\odot$, $a=0.9M$, $d=10$~Mpc, 
$p=3.0$, and $q=0.75$.
However, we adopt $w_{\rm nt}=0.10$ entirely in the jet;
accordingly, both thermal and nonthermal synchrotron components
contribute in emission and absorption.
To elucidate how the spectral shape changes with the matter content,
we adjust the normalization ($B_{\rm p,0}$) of the magnetic-field strength
in each case of different $f_{\rm p}$,
so that the synchrotron fluxes may match in the optically thin regime.

\begin{figure*}
\includegraphics[width=\textwidth, angle=0, scale=1.0]{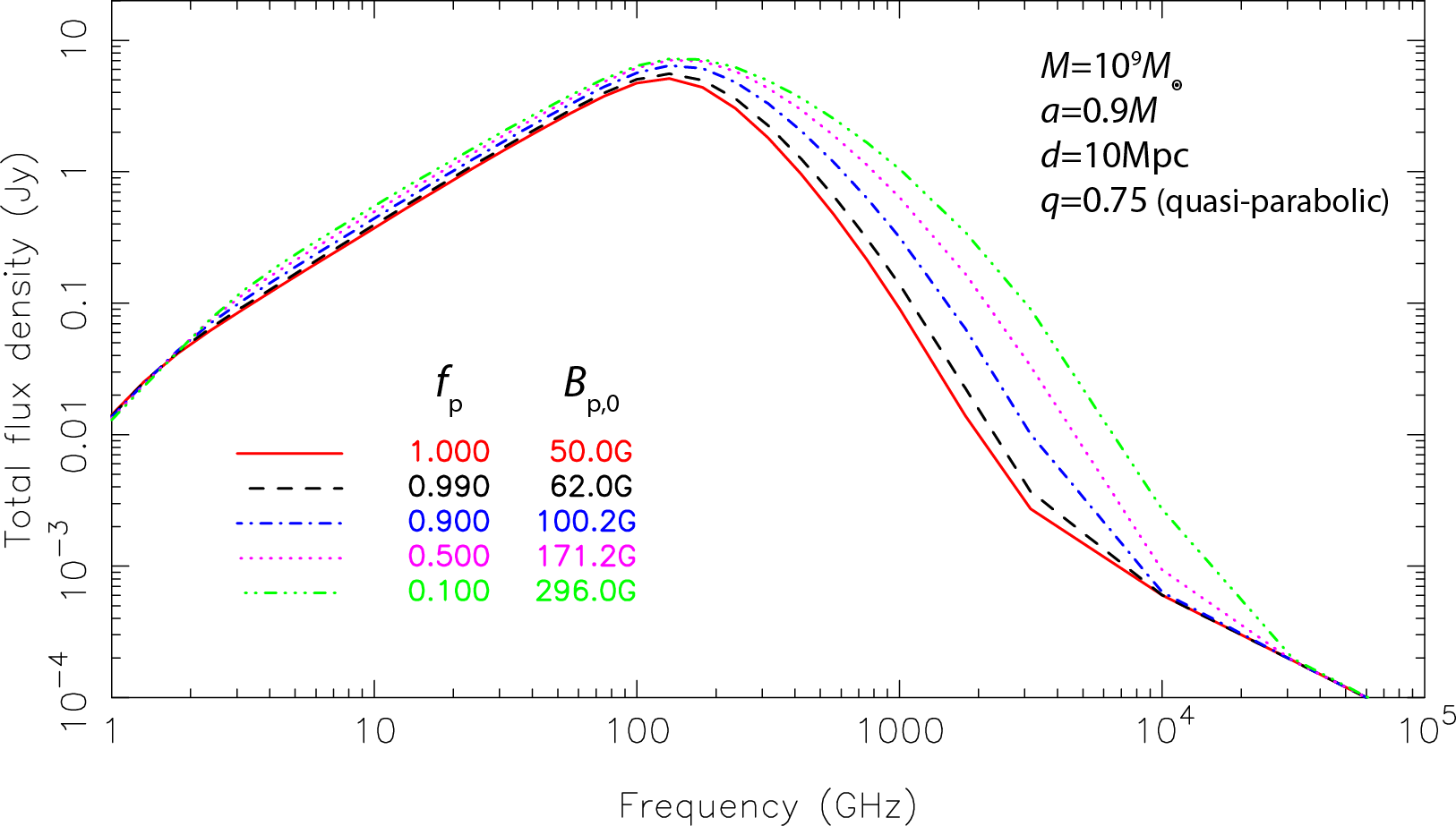}
\caption{
SED of synchrotron process in a BH jet.
The BH is assumed to have mass $M=10^9 M_\odot$
and spin $a=0.9M$, and be located at distance $d=10$~Mpc.
The power-law index of the magnetic-flux function $A_\varphi$ on $r$
is $q=0.75$, which corresponds to a quasi-parabolic flow-line
geometry on the poloidal plane.
The red solid, black dashed, blue dash-dotted, purple dotted, and
green dash-dot-dot-dotted curves show the SED when
the lepton number fraction is
$f_{\rm p}= 1.00$, $0.99$, $0.90$, $0.50$, and $0.10$, respectively.
The magnetic-field strength, $B_{\rm p,0}$ is adjusted as indicated in the figure
so that the flux densities may match in the optically thin regime.
}
    \label{fig:SED_hadron}
\end{figure*}

In figure~\ref{fig:SED_hadron}, we show the SEDs for the five cases.
It shows that the spectrum broadens into the higher frequencies above the peak
when the plasma inertia increases by decreasing $f_{\rm p}$.
This is because the increased magnetic-field strength for smaller $f_{\rm p}$
results in a harder thermal synchrotron emission in the optically thin regime.
In the optically thick regime (i.e., below the peak frequency), 
on the other hand,
the source function tends to be the Planck function,
and hence little depends on the magnetic-field strength;
therefore, there appears little difference in SED.
The peak frequency is approximately given by
$\tau= \alpha_\nu l = 1$,
where $l$ denotes the characteristic length of line of sight
in the jet.
Since $X \equiv \nu_\ast / \nu_{\rm s} \gg 1$ at $\nu=230$~GHz,
$\alpha_\nu^\ast = \ [\delta/(1+z)] \alpha_\nu \propto \exp(-X^{1/3})$.
Thus, the jet becomes more optically thick with increasing $B_{\rm p,0}$
at a fixed $\nu_\ast$,
because $\nu_{\rm s} \propto B \propto B_{\rm p,0}$.
As a result, the spectral shape broadens 
toward higher frequencies above the peak 
when the plasma inertia increases due to hadronic contamination.

\begin{figure*}
\includegraphics[width=\textwidth, angle=0, scale=0.2515]{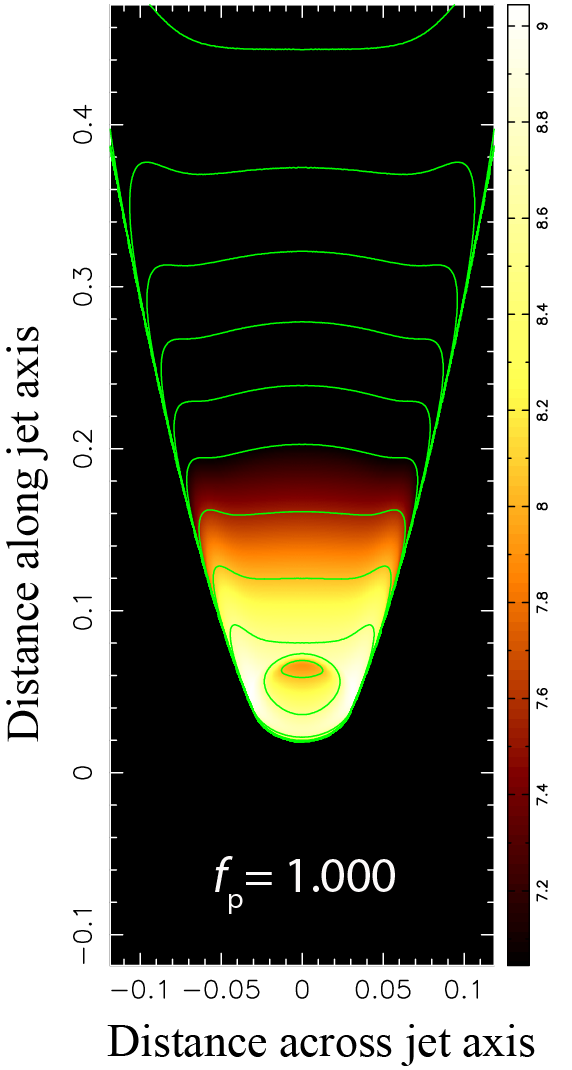}
\includegraphics[width=\textwidth, angle=0, scale=0.22]{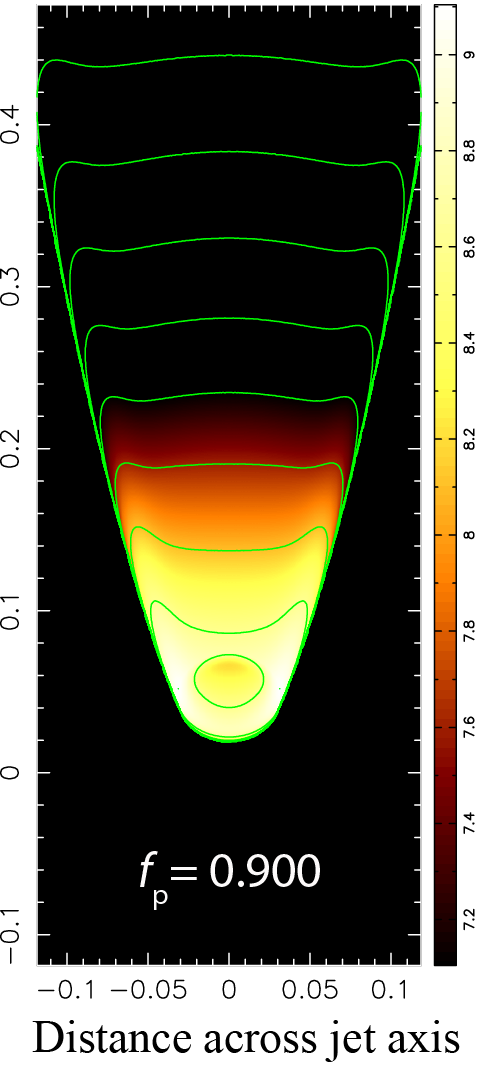}
\includegraphics[width=\textwidth, angle=0, scale=0.22]{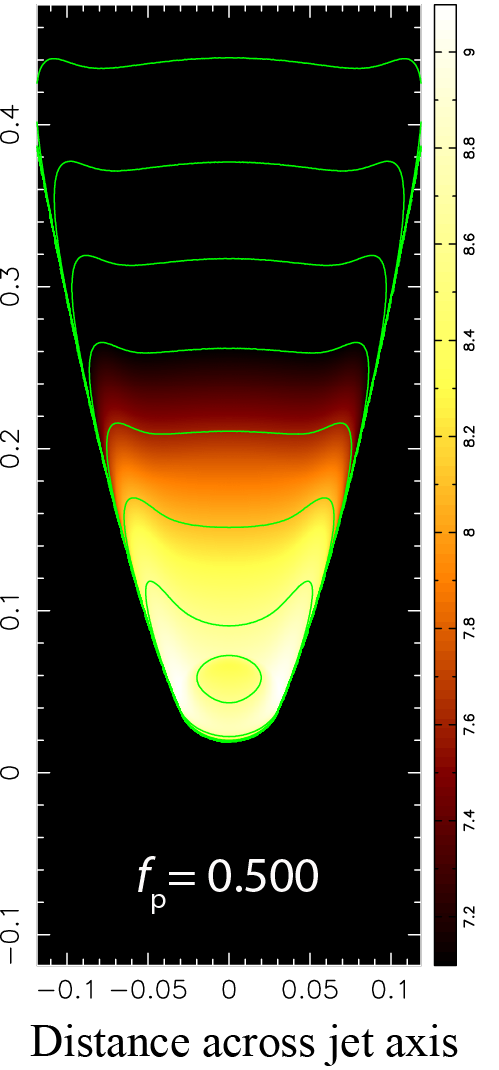}
\includegraphics[width=\textwidth, angle=0, scale=0.22]{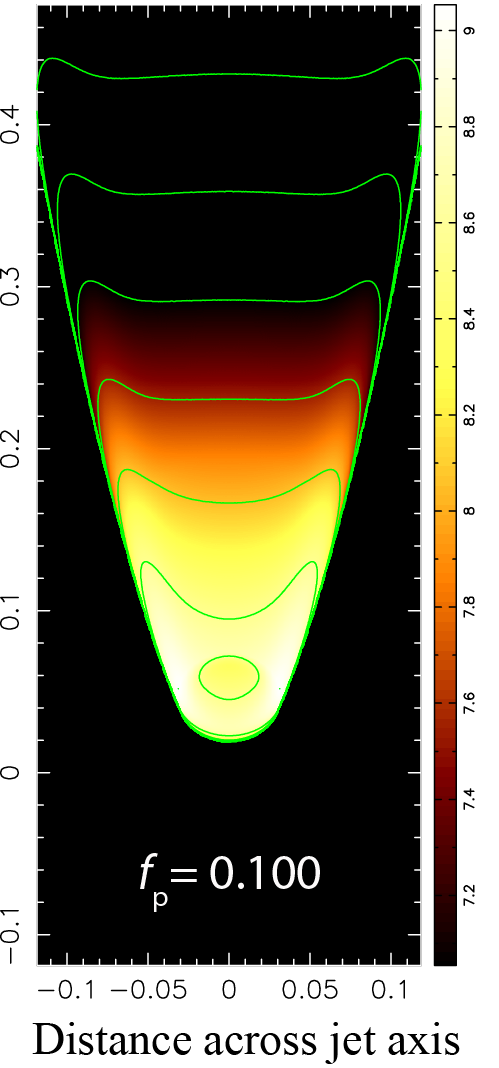}
\caption{
Surface brightness distribution of the base of the approaching jet.
The color is coded logarithmically in $\mu \mbox{Jy mas}^{-2}$ unit.
Thermal leptons are supplied at $r=50 R_{\rm S}$ 
with relativistic temperature, $\Theta_{\rm e}= 10.0$.
The lepton number fraction is
$f_{\rm p}=1.000$, $0.900$, $0.500$, and $0.100$
from left to right.
The peak brightness is $1116 \mbox{Jy mas}^{-2}$,
$1270 \mbox{Jy mas}^{-2}$, 
$1254 \mbox{Jy mas}^{-2}$, and
$1131 \mbox{Jy mas}^{-2}$ respectively.
The color bar covers two decades below the peak brightness
in each panel.
The ring-like structure is enhanced 
for a pair-plasma dominated jet ($f_{\rm p}= 1.0$),
while the limb-brightened structure is enhanced
for a normal-plasma dominated jet ($f_{\rm p}= 0.1$).
}
    \label{fig:map_ring_fp}
\end{figure*}

It is also worth examining how the ring-like structure
changes as a function of the lepton fraction $f_{\rm p}$.
In figure~\ref{fig:map_ring_fp}, we present 
the close-up map of the four cases,
$f_{\rm p}= 1.00$, $0.90$, $0.50$, and $0.10$,
from left to right.
The peak brightness (in $\mbox{Jy mas}^{-2}$) 
is comparable in all the four panels.
In each panel, the color bar covers the brightness 
from the peak value to its $1$~\% value.
It follows that the brightness decreases more gradually
toward the downstream with decreasing $f_{\rm p}$
(i.e., with increasing proton contribution in mass).
As a result, the ring-like structure is blurred
with decreasing $f_{\rm p}$.
On the other hand, 
the limb-brightened structures appear more clearly
at the jet base (in the present case, within the central 0.3 mas),
with decreasing $f_{\rm p}$.

\section{Summary}
\label{sec:summary}
In the {\tt R-JET} code, 
we investigate the jets that are energized by the rotational energy of the BH
via the BZ process.
In this case, the horizon-penetrating magnetic field lines are more tightly
wound in the counter rotational direction with increasing distance
from the rotation axis of the BH.
Accordingly, the BH's rotational energy is preferentially extracted
along the magnetic field lines threading the horizon in the lower latitudes.
We assume that the global magnetic field is axially symmetric with respect to
the spin axis of the BH.
Considering energy conversion from electromagnetic to kinetic ones,
we parameterize the kinetic energy flux at each point with 
the magnetization parameter $\sigma$.
Using this kinetic flux, we constrain the plasma density at each point,
by specifying the bulk Lorentz factor and the internal energy density.
Assuming the fraction of electron-positron pairs in a hadron-contaminated jet,
and assuming the nonthermal fraction of such pairs,
we constrain the spatial distribution of synchrotron-radiating pairs, 
which allows us to compute the emission and absorption coefficients
at each position in the jet.
We then integrate the radiative transfer equation to infer the total intensity,
and hence the brightness distribution of the jet in the celestial plane.

It is confirmed that the {\tt R-JET} code successfully reproduces
the analytic predictions on the turnover frequency as a function of 
the distance $r$ from the BH for thermal pairs,
and on the coreshift as a function of the photon frequency
for nonthermal pairs.
Then we apply the {\tt R-JET} code to typical jet parameters
and demonstrate that the jets naturally exhibit limb-brightened structure
by virtue of the angle-dependent energy extraction from the BH via the BZ process.
We also show that a ring-like bright structure will be detected by VLBI technique,
if hot plasmas are supplied in the jet launching region,
and if we observe the jet nearly face-on.

In our subsequent papers,
we will apply the {\tt R-JET} code to individual AGN jets
and examine the SED, the coreshift, limb-brightened and ring-like structures, 
adopting the parameter sets that are constrained by observations.
After incorporating this post-processing, {\tt R-JET} code 
into our future fluid code, 
we intend to make the complete package publicly available.

\begin{acknowledgments}
The authors acknowledge grant support for the CompAS group under Theory from the Institute of Astronomy and Astrophysics, Academia Sinica (ASIAA), and the National Science and Technology Council (NSTC) in Taiwan through grants 112-2112-M-001-030- and 113-2112-M-001-008. The authors acknowledge the access to high-performance facilities (TIARA cluster and storage) in ASIAA and thank the National Center for High-performance Computing (NCHC) of National Applied Research Laboratories (NARLabs) in Taiwan for providing computational and storage resources. 
This work utilized tools (BIWA GRPIC code) developed and maintained 
by the ASIAA CompAS group.
This research has made use of SAO/NASA Astrophysics Data System.
\end{acknowledgments}

\appendix

\section{Poloidal magnetic field}
\label{sec:app_Bp}
In this appendix, 
we present the expressions of a magnetic field in the poloidal plane.
Magnetic field is defined by the Maxwell tensor, and becomes
$B^r= \epsilon_t{}^r{}_{\theta \varphi} F^{\theta \varphi}$ and
$B^\theta= \epsilon_t{}^\theta{}_{\varphi r} F^{\varphi r}$
in the Boyer-Lindquist coordinates,
where $\epsilon_{\mu\nu\rho\sigma}$ denotes
the Levi-Civita symbol, 
$F^{\mu\nu}$ does (the contravariant component of) the Faraday tensor,
and $g_{\mu\nu}$ the metric tensor
\citep[e.g., eqs.~(1)--(3) of ][]{Hirotani:2023:ApJ}.
Substituting the definition of $\Omega_F$
(namely, the angular-frequency of the rotating magnetic field), 
$F_{\theta t}= -\Omega_F F_{\theta\varphi}$ and
$F_{r t}= \Omega F_{\varphi r}$,
we obtain
\citep{Bekenstein:1978:PhRvD,came86b,takahashi:1990ApJ}
\begin{equation}
  B^r= \frac{-g_{tt}-g_{t \varphi}\Omega_F}{\sqrt{-g}} F_{\theta\varphi},
  \label{eq:app_Br_2}
\end{equation}
\begin{equation}
  B^\theta= \frac{-g_{tt}-g_{t \varphi}\Omega_F}{\sqrt{-g}} F_{\varphi r},
  \label{eq:app_Bth_2}
\end{equation}
where $F_{\theta\varphi}= \partial_\theta A_\varphi$ and
$F_{\varphi r}= -\partial_r A_\varphi$
in an axisymmetric magnetosphere.

In the jet region, $r \gg M = GM c^2$,
substituting equation~(\ref{eq:A3}) into these two equations,
we obtain equation~(\ref{eq:Bp}),
where $B_{p,0} \equiv A_{\rm max}/R_{\rm S}{}^2$ and
$q=0$ is assumed at $r= R_{\rm S}$.

Near the horizon, $\Delta \approx 0$,
we regard $q$ is constant for both $r$ and $\theta$.
Then, equations~(\ref{eq:app_Br_2})--(\ref{eq:app_Bth_2}) 
and equation~(2) of H24 give equation~(\ref{eq:Bp}).
It is worth noting that equation~(\ref{eq:Bp}) is correct
in a stationary and axisymmetirc magnetosphere 
not only near the horizon but also far from it,
as long as $q$ is constant.

\bibliography{hirotani.bib}{}
\bibliographystyle{aasjournal}

\end{document}